\newcommand{\gev}{\, {\rm GeV}}
\newcommand{\beq}{\begin{equation}}
\newcommand{\eeq}{\end{equation}}
\newcommand{\bea}{\begin{eqnarray}}
\newcommand{\eea}{\end{eqnarray}}

\documentclass[aps,prl,reprint,twocolumn,preprintnumbers,floatfix,nofootinbib]{revtex4-1}

\usepackage{graphicx}
\usepackage{epstopdf}
\usepackage{mathrsfs}
\usepackage{amssymb}
\usepackage{verbatim}
\usepackage{color}
\usepackage{multirow}
\usepackage{amsmath}
\usepackage{subfig}
\usepackage{slashed} 
\usepackage{float}
\usepackage[normalem]{ulem}
\usepackage{sidecap}
\usepackage[utf8]{inputenc}

\usepackage{hyperref}

\def\sm{\text{\tiny{SM}}}

\def\rh{\text{\tiny{RH}}}

\def\a{\alpha}
\def\b{\beta}

\def\m{\mu}
\def\n{\nu}

\def\vp{\varphi}

\def\G{\Gamma}
\def\L{\Lambda}
\def\O{\Omega}

\def\gsm{g_{\text{\tiny{SM}}}}
\def\gdm{g_{\text{\tiny{DM}}}}

\def\MP{M_P}
\def\mh{m_{\tilde{h}}}
\def\Gh{\Gamma_{\tilde h}}

\def\to{\rightarrow}

\def\M{\mathcal{M}}

\begin{document}\sloppy  

%
%

\preprint{LPT--Orsay 18-13}
\preprint{UMN--TH--3710/18}
\preprint{FTPI--MINN--18/02}
\preprint{PI/UAN--2018--619FT}

\vspace*{1mm}

\title{Spin-2 Portal Dark Matter}
\author{Nicol\'as Bernal$^{a}$}
\email{nicolas.bernal@uan.edu.co}
\author{Ma\'ira Dutra$^{b}$}
\email{maira.dutra@th.u-psud.fr}
\author{Yann Mambrini$^{b}$}
\email{yann.mambrini@th.u-psud.fr}
\author{Keith Olive$^{c,d}$}
\email{olive@umn.edu}
\author{Marco Peloso$^{c}$}
\email{peloso@umn.edu}
\author{Mathias Pierre$^{b}$}
\email{mathias.pierre@th.u-psud.fr}
\vspace{0.5cm}
\affiliation{
${}^a$ 
Centro de Investigaciones, Universidad Antonio Nariño\\
Carrera 3 Este \# 47A-15, Bogotá, Colombia}
\affiliation{
${}^b$ 
Laboratoire de Physique Théorique (UMR8627), CNRS, Univ. Paris-Sud, Université Paris-Saclay, 91405 Orsay, France}
\affiliation{$^c$School of
 Physics and Astronomy, University of Minnesota, Minneapolis, MN 55455,
 USA} 
\affiliation{$^d$William I. Fine Theoretical Physics Institute, School of
 Physics and Astronomy, University of Minnesota, Minneapolis, MN 55455,
 USA}

\begin{abstract} 
We generalize models invoking a spin-2 particle as a mediator between the dark sector and the Standard Model. We show that a massive spin-2 messenger can efficiently play the role of a portal between the two sectors. The dark matter is then produced via a freeze-in mechanism during the reheating epoch. In a large part of the parameter space, production through the exchange of a massive spin-2 mediator dominates over processes 
involving a graviton with Planck suppressed couplings. We perform a systematic analysis of such models for different values of the spin-2 mass relative to the maximum and the final temperature attained at reheating.   
 
\end{abstract}

\maketitle

\maketitle


\setcounter{equation}{0}



\section{Introduction}

Although there is a large amount of indirect evidence for the presence of dark matter in the Universe from astrophysical observations, its precise nature remains elusive. 
Moreover, although there has been an impressive increased sensitivity in direct and indirect 
detection searches, no signal has been confirmed so far. Indeed, the recent results of direct detection experiments like LUX \cite{Akerib:2016vxi}, PANDAX-II \cite{Cui:2017nnn} or XENON1T \cite{Aprile:2017iyp} constrain a large part of the WIMP parameter space, and are close to
excluding the simplest extensions of the Standard Model (SM) such as the Higgs-portal model
\cite{Silveira:1985rk,McDonald:1993ex,Burgess:2000yq,Davoudiasl:2004be,hportal,Han:2015hda,Mambrini:2016dca}, the Z-portal \cite{zportal,Escudero:2016gzx} or even the Z'-portal \cite{zpportal} 
(see \cite{Arcadi:2017kky} for a recent review on the subject). 

There are, however, many alternatives where the dark matter is 
secluded from SM particles by intermediate mass scale messengers.
Indeed, intermediate scales are naturally present in many well-motivated extensions of 
the SM, including Grand Unified Theories, models with a see-saw mechanism, string constructions, inflation and reheating or leptogenesis. 
In all of these frameworks, the presence of an intermediate mass scale generates a heavy particle
spectrum which can in principle mediate interactions between a possible dark sector and the SM.
Some specific examples of these frameworks are Grand Unified SO(10) models \cite{so10} or high-scale supersymmetry \cite{Benakli:2017whb,gravitino}, where dark matter candidates respecting the Planck/WMAP constraints \cite{Hinshaw:2012aka,Ade:2015xua} are present.
The effective superweak coupling generated by 
the exchange of an intermediate mass or even a superheavy mediator much heavier than the reheating temperature after inflation, $T_\text{RH}$,  allows for
the production of dark matter directly from the thermal bath in the same way that gravitinos are produced during reheating by Planck suppressed operators \cite{Nanopoulos:1983up,Ellis:1983ew,Khlopov:1984pf}. Qualitatively similar results 
can be obtained with lighter mediators and small couplings as in the freeze-in 
mechanism (see \cite{fimp} for a recent review on the subject).

Often it is sufficient to approximate the details of 
particle production during reheating with the instantaneous reheating approximation.
Namely, that all particle production occurs at the end of the reheating
process characterized by the reheating temperature $T_\text{RH}$.
However, depending on the specific production process, 
the instantaneous approximation may or may not be a good approximation.
For example, if we parameterize the production cross section by
\begin{equation}
\left\langle \sigma v \right\rangle \simeq \frac{T^n}{M^{n+2}} \, , 
\label{sv}
\end{equation}
where $M$ is some energy scale and $T$ the temperature, this approximation has been shown to be reasonable for $n< 6$ \cite{Chung:1998rq,Giudice:2000ex,Kolb:2003ke,Garcia:2017tuj}.
However, for $n \ge 6$, the production rate is sensitive to the maximum temperature during the reheating process \cite{Ellis:2015jpg,Garcia:2017tuj} as we describe in 
more detail below. Thus the detailed mechanism for dark matter
production during reheating will in general be sensitive to 
the form of the coupling of the dark sector to the SM.

In the case of the gravitino, as noted above, from the strict observational point of view, dark matter only requires gravitational coupling. That is, communication between the two sectors (dark and SM) is mediated only through gravity, which couples to the energy--momentum tensor
of the standard model and dark matter. Because this coupling is fixed by the equivalence principle, the only free parameter in this model is the mass of the dark matter.
It has been shown in \cite{Garny:2017kha} that independent of the nature of dark matter, there exists a possibility to populate the relic abundance through a freeze-in mechanism via the exchange of a massless spin-2 graviton. 

In this work we generalize this to the case in which the exchanged spin-2 particle is massive. 
We borrow concepts from the theory of massive gravity. The Lorentz invariant linear theory of a massive graviton was formulated by Fierz and Pauli in \cite{Fierz:1939ix}, where it was shown that only one specific choice for the mass term is free from singularities. At the linear level, the massive graviton has 5 polarizations, as expected for a massive spin-2 particle. It was then shown that a generic nonlinear completion introduces a sixth state, that is a ghost \cite{Boulware:1973my}. Ref. \cite{deRham:2010kj} provided a nonlinear construction that is ghost free.~\footnote{In fact, such a model was already formulated in \cite{GrootNibbelink:2004hg,Nibbelink:2006sz}, using the vielbein formalism. It was noted in these works that the ghost does not appear in the scalar sector of the theory. A conclusive proof that the ghost is absent in the full theory was later obtained in \cite{Hassan:2011hr}, also using the vielbein formulation.} We are not interested here in the nonlinear self-interactions of the massive spin-2 field, so we will simply employ the Fierz and Pauli  linear term, implicitly assuming a ghost-free nonlinear completion.

The coupling of the spin-2 mediator is expected to be universal,
but it might couple more strongly to the SM (and the dark sector) than a 
Planck suppressed gravitational coupling. Thus we consider here,
a massive spin-2 mediator coupling via the energy momentum tensor
but with an intermediate mass scale, thus enhancing its couplings 
relative to gravity. We generalize spin-2 couplings to dark matter and study the production mechanism of dark matter through a massive spin-2 portal. 

The paper is organized as follows. In the next section, we lay out the model which includes a massive spin-2 mediator which is coupled to 
both the dark matter sector and the SM. In section III,
we discuss our computation of the relic dark matter abundance
and our results are given in section IV where we consider separately
the cases of heavy and light mediators and discuss the impact of dropping the instantaneous reheating approximation and our conclusions are given in section V.

\section{II. The model}

The model we consider is a relatively minimal 
extension of the SM which includes (in addition to the massless graviton $h_{\mu \nu}$) 
a dark matter candidate $X$, and a massive spin-2 mediator $\tilde h_{\mu \nu}$ and is described by
the Lagrangian
\beq
{\cal L} = {\cal L}_\text{SM} + {\cal L}_\text{DM} + {\cal L}_\text{EH} + {\cal L}_{\tilde h}+{\cal L}^1_\text{int}
+ {\cal L}^2_\text{int} \, ,
\eeq
with ${\cal L}_\text{SM}$ (${\cal L}_\text{DM}$) the standard model 
(dark matter) Lagrangian. ${\cal L}_\text{EH}$ is the Einstein-Hilbert sector which contains the kinetic terms of the massless graviton obtained after expanding the metric around flat space, $g_{\mu \nu} \simeq \eta_{\mu \nu} + h_{\mu \nu}/M_P$. ${\cal L}_{\tilde h}$ is the ghost-free Fierz-Pauli Lagrangian which contains the kinetic and mass terms for the massive spin-2 field.  The final two terms,
${\cal L}^i_\text{int}$ are the interaction Lagrangians with the massless graviton 
$h_{\mu \nu}$ ($i=1$) and the massive spin-2 mediator $\tilde h_{\mu \nu}$ ($i=2$)
that can be written, from the equivalence principle:
\beq
{\cal L}^1_\text{int} = \frac{1}{2 M_{P}}h_{\mu \nu}~(T^{\mu \nu}_\text{SM}+ T^{\mu \nu}_\text{X})
\eeq
\beq
{\cal L}^2_\text{int} =  \frac{1}{\Lambda} \tilde h_{\mu \nu}~ 
(\gsm T^{\mu \nu}_\text{SM}+ \gdm T^{\mu \nu}_\text{X})
\eeq
where $M_P$ is the reduced Planck mass $M_P \simeq 2.4\times 10^{18}$ GeV, and $\Lambda \lesssim M_P$ is an
intermediate scale and governs the strength of the new spin-2 interaction. The couplings, $\gsm$ ($\gdm$)
 of the messenger to the standard model (dark matter)
 allow us to distinguish interactions between the two sectors. 
 Of course only 2 of the three parameters ($\Lambda, \gsm,
 \gdm$) are independent. 
 
 The form of the stress-energy tensor of a field,
$T^a_{\mu \nu}$ depends on its spin $a=0,1/2,1$.\footnote{We assume real scalars and Dirac 
fermions throughout our work.} In general, we can write
\bea
T^0_{\mu \nu} &=& \frac{1}{2} \left( \partial_\mu \phi~\partial_\nu \phi + \partial_\nu \phi ~\partial_\mu \phi -g_{\mu \nu} \partial^\alpha \phi ~\partial_\alpha \phi \right) \,, \nonumber\\
T^{1/2}_{\mu \nu} &=& \frac{i}{4}
\bar \psi \left( \gamma_\mu \partial_\nu + \gamma_\nu \partial_\mu \right) \psi
-\frac{i}{4} \left( \partial_\mu \bar \psi \gamma_\nu + \partial_\nu \bar \psi \gamma_\mu \right)\psi \,, 
\nonumber\\
T^{1}_{\mu \nu} & = & \frac{1}{2} \left[ F_\mu^\alpha F_{\nu \alpha} + F_\nu^\alpha F_{\mu \alpha} - \frac{1}{2} g_{\mu \nu} F^{\alpha \beta} F_{\alpha \beta} \right] \,.
\eea
The amplitudes relevant for the computation of the processes $\text{SM}^a(p_1)+\text{SM}^a(p_2) \rightarrow \text{DM}^b(p_3)+\text{DM}^b(p_4)$ can be parametrized by 
\begin{equation}
\mathcal{M}^{ab} \propto \sum_{i=1,2} \langle p_1^a p_2^a | \mathcal{L}_\text{int}^i | p_3^b p_4^b \rangle \propto \sum_{i=1,2} M_{\mu \nu}^a \Pi^{\mu \nu \rho \sigma}_i M_{\rho \sigma}^b \;, 
\end{equation}
where $b$ denotes the spin of the DM involved in the process and $b=0,1/2,1$. $\Pi^{\mu \nu \rho \sigma}_i$ denotes the propagators of the graviton ($i=1$) and massive spin-2 ($i=2$) which are given in the Appendix.
The partial amplitudes, $M_{\mu \nu}^a$, can be expressed as
\bea 
M_{\mu \nu}^0 &=& \frac{1}{2}(p_{1\mu} p_{2\nu} + p_{1\nu} p_{2\mu} - g_{\mu \nu}p_1.p_2) \,, \nonumber\\ 
M_{\mu \nu}^{1/2} &=&  \frac{1}{4} {\bar v}(p_2) \left[ \gamma_\mu (p_1-p_2)_\nu + \gamma_\nu (p_1-p_2)_\mu \right] u(p_1) \;,  \nonumber\\ 
M_{\mu \nu}^{1} &=&  \frac{1}{2}\Bigg[ 
\epsilon_2^*.\epsilon_1(p_{1\mu} p_{2\nu}+ p_{1 \nu} p_{2 \mu})- \epsilon_2^*.p_1
(p_{1 \mu} \epsilon_{1 \nu} + \epsilon_{1 \mu} p_{2 \nu}) \nonumber
\\
&&
-\epsilon_1.p_2 (p_{1 \nu} \epsilon^*_{2 \mu} + p_{1 \mu} \epsilon^*_{2 \nu})
+p_1.p_2(\epsilon_{1 \mu} \epsilon^*_{2 \nu}+ \epsilon_{1 \nu} \epsilon^*_{2 \mu})
\nonumber
\\
&&
+\eta_{\mu \nu}(\epsilon_2^*.p_1 \epsilon_1.p_2 - p_1.p_2~ \epsilon^*_2.\epsilon_1)
\Bigg] \,, 
\eea
with similar expressions in terms of the dark matter momenta, $p_3, p_4$. 

The total amplitude squared implied in the processes SM SM $\rightarrow$ DM DM will be a sum of the three contributions, weighted by the standard model content in fields:
\beq
|{\cal M}|^2= 4 |{\cal M}^0|^2 + 45 |{\cal M}^{1/2}|^2 + 12 |{\cal M}^1|^2.
\eeq
Further details regarding these amplitudes are found in the Appendix.

\section{III. The relic abundance}

As noted in the introduction, reheating after inflation is often assumed to occur instantaneously, on a timescale given by the inflaton decay rate $\Gamma_\phi$. This results in a thermal bath of initial temperature (see \cite{Ellis:2015jpg,Garcia:2017tuj} for a detailed discussion) 
\begin{equation}
T_{\rm RH} = \left( \frac{40}{g_{\rm RH} \, \pi^2} \right)^{1/4} \left( \frac{\Gamma_\phi \, M_p}{c} \right)^{1/2} \,, 
\end{equation}
where $g_{\rm RH}$ is the number of effective degrees of freedom in the thermal bath of temperature $T_{\rm RH}$, and where $c$ is an order one parameter that depends on when precisely reheating is assumed to take place (setting $\Gamma_\phi^{-1}$ equal to the reheating time, $\Gamma_\phi^{-1} = t_{\rm RH}$, leads to $c=1$; 
setting it instead equal to the Hubble rate, $\Gamma_\phi = H \left( t_{\rm RH} \right)$, leads to $c=\frac{2}{3}$). Numerical solutions to particle yields during reheating agree with the
instantaneous approximation if $c \approx 1.2$ \cite{Pradler:2006hh,Rychkov:2007uq,Ellis:2015jpg}.

In reality reheating is a finite-duration process. The inflaton decay products thermalize on a much shorter timescale than $\Gamma_\phi^{-1}$, so it is appropriate to assume the co-existence of a decaying inflaton, and of a thermal bath arising from the decay. In this context, the reheating temperature is conventionally defined as the temperature of the thermal bath when it starts to dominate over the inflaton. However, the thermal bath reaches its maximum temperature while still subdominant to the inflaton. One finds (see for instance Ref~\cite{Ellis:2015jpg}) 
\begin{equation}
T_{\rm max} \simeq 0.5 \left( \frac{m_\phi}{\Gamma_\phi} \right)^{1/4} \, T_{\rm RH} \,, 
\end{equation}
where $m_\phi$ is the inflaton mass. Perturbativity requires $\Gamma_\phi < m_\phi$, and it is typical to have $\Gamma_\phi \ll m_\phi$. Therefore the maximum temperature of the thermal bath can be many orders of magnitude greater than $T_{\rm RH}$. 

Particle production at temperatures $T > T_{RH}$ can be significant.  There are two reasons, in fact,  why assuming an instantaneous reheating can lead to a significant underestimation of the dark matter abundance. The first case, which has extensively been pointed out in the literature, occurs when the (thermally averaged) dark matter production cross section times velocity given in Eq. (\ref{sv})  has a strong temperature dependence. 
Such a relation applies for instance when the dark matter is produced from quanta in the thermal bath by the exchange of a heavy mediator of mass $M \gg T$. If $n< 6$ in this relation, the dark matter abundance is mostly produced at the end of reheating, when $T \simeq T_{\rm RH}$. In contrast, the quanta produced at $T \simeq T_{\rm max}$ dominate the final abundance if $n > 6$. For $n=6$, particles produced at any temperature equally contribute to the final abundance. This results in a logarithmic $\ln \frac{T_{\rm max}}{T_{\rm RH}}$ enhancement of the abundance with respect to  the naive estimate based on instantaneous reheating. As shown in \cite{Garcia:2017tuj}, $n=6$ is obtained in the case of gravitino dark matter in high scale supersymmetry models. As we show in the present work, this also applies to the cases in which the mediator is a heavy spin-2 particle. While we do not consider it here, the same strong temperature dependence is found if the mediator is a pseudo-scalar particle, and both the dark matter and the quanta in the thermal bath are spin-1 particles. 

A second reason why the instantaneous reheating case can lead to a significant underestimation of the dark matter abundance, and which we explore in the present work, is if the mass $M$ of the mediator is between $T_{\rm RH}$ and $T_{\rm max}$. In this case, it is possible that the final dark matter abundance is due to the quanta produced on resonance, taking place at $T \simeq M$. 
The resonance is missed if one simply assumes that the thermal bath is instantaneously formed with $T = T_{\rm RH}$. 

The numerical analysis in this work takes into account the total set of Boltzmann equations for the time evolution of a system whose energy density is in the form of unstable massive particles $\phi$ (the inflaton for instance), stable massive DM particles $X$, and radiation $R$ (ie SM particles).
For the exact computation, 
we assumed that $\phi$ decays into radiation with a rate $\Gamma_\phi$, and that the DM particles are created and annihilate into radiation with a thermal-averaged cross section times velocity $\langle\sigma v\rangle$.
The corresponding energy and number densities satisfy the differential equations~\cite{Chung:1998rq, Giudice:2000ex}
\bea
\frac{\text{d}n_X}{dt}&=&-3H\,n_X-\langle\sigma v\rangle\left[n_X^2-(n_X^\text{eq})^2\right]\,, \nonumber \\
\frac{\text{d}\rho_R}{dt}&=&-4H\,\rho_R+\Gamma_\phi\,\rho_\phi+2\langle\sigma v\rangle\langle E_X\rangle\left[n_X^2-(n_X^\text{eq})^2\right]\,, \nonumber \\
\frac{\text{d}\rho_\phi}{dt}&=&-3H\,\rho_\phi-\Gamma_\phi\,\rho_\phi\,.
\label{Eq:setboltzmann}
\eea
We assumed that each $X$ has energy $\langle E_X\rangle\simeq\sqrt{m_X^2+ 9T^2}$ and the factor $\langle E_X\rangle$ is the average energy released per $X$ pair annihilation.
The Hubble expansion parameter $H$ is given by $H^2=\frac{1}{3M_P^2}(\rho_\phi+\rho_R+\rho_X)$.

Thermalization of the SM radiation produced by inflaton decays is rapid \cite{Davidson:2000er,Harigaya:2013vwa,Mukaida:2015ria,Ellis:2015jpg}. As such, we can define a 
radiation temperature, 
\beq\label{inst_tempe}
T = \left(\frac{30\rho_{R}}{\pi^2 g(T)}\right)^{1/4}\,,
\eeq
and $T_\text{\scriptsize{max}}$ corresponds to the maximum temperature 
attained during the reheating process.

When the dark matter number density is far below its equilibrium abundance (and when inflaton decays do not directly produce $X$) the dark matter density $n_X$ is given by the approximate Boltzmann equation
\beq
\frac{\text{d} n_X}{\text{d}t} = -3 H(t) n_X + (n_X^\text{eq})^2 \langle \sigma v \rangle  \,, 
\eeq
where $H \left( t \right)$ is the time dependent Hubble rate, and where $n_X^\text{eq}$ is the number density that the dark-matter would have in thermal equilibrium. This relation assumes that the dark matter is produced through $2 \rightarrow 2$ processes from quanta in the thermal bath, and that the dark matter abundance is well below its thermal equilibrium value, which is the case in the models considered here. 

This relation can be rewritten as 
\beq
\frac{\text{d} Y_X}{\text{d}T} = - \frac{R(T)}{H~T~s} \,, 
\label{Eq:y}
\eeq
where $Y_X=n_X/s$ is the dark matter yield, 
$s=\frac{2 \pi^2}{45} g_s(T) T^3$ is the 
entropy density in the thermal bath with $g_s(T)$ effective number of degrees of freedom. The production rate, $R(T)= (n_X^\text{eq})^2 \langle \sigma v \rangle$ for the 1 + 2 $\rightarrow$ 3 + 4 process is obtained from
\beq
R(T) = \int f_1 f_2 \frac{E_1 E_2 \text{d}E_1 \text{d}E_2 ~\text{d}\cos \theta_{12}}{1024 \pi^6} \int |{\cal M}|^2 \text{d} \Omega_{13} \,,  
\label{Eq:rt}
\eeq
where $f_1$ and $f_2$ are the distribution functions of the initial (SM) particles.

The two processes we consider contributing to the relic abundance are the exchange of the (massless) graviton $h_{\mu \nu}$, with Planck mass suppressed couplings, and the exchange of the massive spin-2 mediator $\tilde h_{\mu \nu}$. Different results are obtained in the heavy mediator ($m_{\tilde h} > T_{RH}$) and light mediator ($m_{\tilde h} < T_{RH}$) cases which are discussed separately below.

\section{IV. Results}

As noted above, all of the results in this paper are obtained via a numerical calculation using the complete set of the Boltzmann equations (\ref{Eq:setboltzmann}), and are not based on the instantaneous reheating approximation. However, it is useful to give approximate solutions in order to perform simple analytical estimates, as the difference between the instantaneous and the non-instantaneous reheating is often an overall multiplicative factor \cite{Garcia:2017tuj}. 
In the Appendix, we derive in detail the computation of the relic abundance. We obtained the following expression for the relic density in the instantaneous approximation:
\begin{widetext}
\begin{equation}
\begin{split}
\frac{\O h^2_\rh}{0.1} \approx \left( \frac{m_X}{1 \gev} \right) \Big\{ & 8 \times 10^{-17} \Big(\frac{\a}{\alpha^0 }\Big) \Big(\frac{T_\rh}{10^{12}\gev}\Big)^3 + \\ 
& + \Theta[T_\rh - \mh] \Big[ 4 \times 10^{-6} \Big(\frac{\b_1}{\beta_1^0 }\Big) \Big(\frac{T_\rh}{10^{12}\gev}\Big)^3 \Big(\frac{10^{16}\gev}{\L}\Big)^4\\
& \hspace{2.6cm} + 2 \Big(\frac{\b_2}{\beta_2^0 }\Big) \Big(\frac{\mh}{10^{10}\gev}\Big) \Big(\frac{10^{16}\gev}{\L}\Big)^2 \Big]\\
& +  \Theta[\mh - T_\rh] \Big[ 70  \Big(\frac{\b_3}{\beta_3^0}\Big) \Big(\frac{T_\rh}{10^{12}\gev}\Big)^7  \Big(\frac{10^{16}\gev}{\L}\Big)^4  \Big(\frac{10^{11}\gev}{\mh}\Big)^4  \Big] \Big\} \,, 
\label{Eq:omega}
\end{split}
\end{equation}
\end{widetext}
where we fixed $\gsm=\gdm=1$, $g_s=100$ and $\Theta$ is the Heaviside step function.
The first term in this expression is due to ordinary graviton exchange in the processes producing dark matter. We see that this contribution is completely negligible unless the dark matter is very heavy. The following terms are due to the exchange of a massive spin-2 state in three different mass regimes, namely mass greater, smaller but comparable, and much smaller than the reheating temperature. The numerical factors $\a, \b_1, \b_2$ and $\b_3$ are normalized by their values for the case of a scalar dark matter candidate and would change by a factor of one order of magnitude if the dark matter is a fermion or a vector. The superscript of the coupling coefficients denotes the 
spin of the dark matter candidate and values of these coefficients are tabulated in a
table given in the Appendix. Note that the parametric dependence in the above expressions for $\Omega h^2$ can be seen directly from the rates with 
$\Omega h^2 \sim R/nH \sim R M_P/T^5$.

Overall, we can distinguish 4 regimes corresponding to 4 different process leading to the production of a sufficient abundance of dark matter.

\clearpage

\begin{itemize}
\item{{\it Super-heavy mediator regime}, or decoupling regime ($\mh \gg T_\text{RH}$):
When $\mh \gtrsim 3000 (\beta_3 \alpha^0/\beta_3^0 \alpha)^{1/4} (10^{16} {\rm GeV}/\Lambda) T_\text{RH}$, the dominant production mode is through the exchange of a massless graviton. For very large masses, the rate mediated by the spin-2 propagator is suppressed even relative to the Planck suppressed rate mediated by gravity. In this regime, the first term of Eq.~(\ref{Eq:omega}) dominates. The production rate in such regime is very sensitive to the temperature of the thermal bath and is proportional to $R(T) \propto \frac{T^8}{M_P^4}$. 
We have collected all of the rates in the Appendix for the separate cases of scalar,
fermionic, and vector dark matter.}

\item{{\it Heavy mediator regime} ($\mh > T_\text{RH}$): When the condition for the super-heavy mediator regime above is not satisfied, yet the mediator mass still exceeds the reheating temperature, the dominant mode is massive spin-2 mediator with rate given by the final term (proportional to $\beta_3$) in Eq. (\ref{Eq:omega}). Despite the massive mediator, the coupling and hence the rate are enhanced over the gravitational rate by the fact that $\Lambda < M_P$. The rate is in this case $highly$ dependent on the temperature and is proportional to $R(T) \propto \frac{T^{12}}{\Lambda^4 \mh^4}$.}

\item{{\it Narrow Width Approximation} (NWA) ($\mh \lesssim T_\text{RH}$): this regime dominates when the temperature of the thermal bath approaches the mediator mass $\mh$. $\tilde h_{\mu \nu}$ is then produced on-shell in resonance, and if the width $\Gamma_{\tilde h}$ is sufficiently small, this process will dominate. 
The width is calculated in the Appendix and scales as $\Gamma_{\tilde h} \sim \mh^3/\Lambda^2$. 
The rate is mildly dependent on the temperature and is proportional to $R(T)\propto \frac{\mh^9}{\Lambda^4\Gamma_{\tilde h}} \frac{T}{\mh} ~K_1\left(\frac{\mh}{T}\right) 
\propto \frac{\mh^6}{\Lambda^2} \frac{T}{\mh} K_1 \left(  \frac{\mh}{T}\right)$ (see the Appendix for the details). For $T \sim \mh$ we obtain the term proportional to $\b_2$ in 
Eq. (\ref{Eq:omega}).} 

\item{{\it Light mediator regime} ($\mh \ll T_\text{RH}$):
this regime is very similar to the well studied case of a light gravitino. Indeed, the behavior and couplings are exactly the same except the coupling, proportional to $1 / \Lambda$ and not $1/ M_P$, is much larger. This regime is the one for which the particle production is greatest and the rate is proportional to $R(T)\propto \frac{T^8}{\Lambda^4}$ }.
\end{itemize}

\begin{figure}[h!]
\centering
\includegraphics[width=0.47\textwidth]{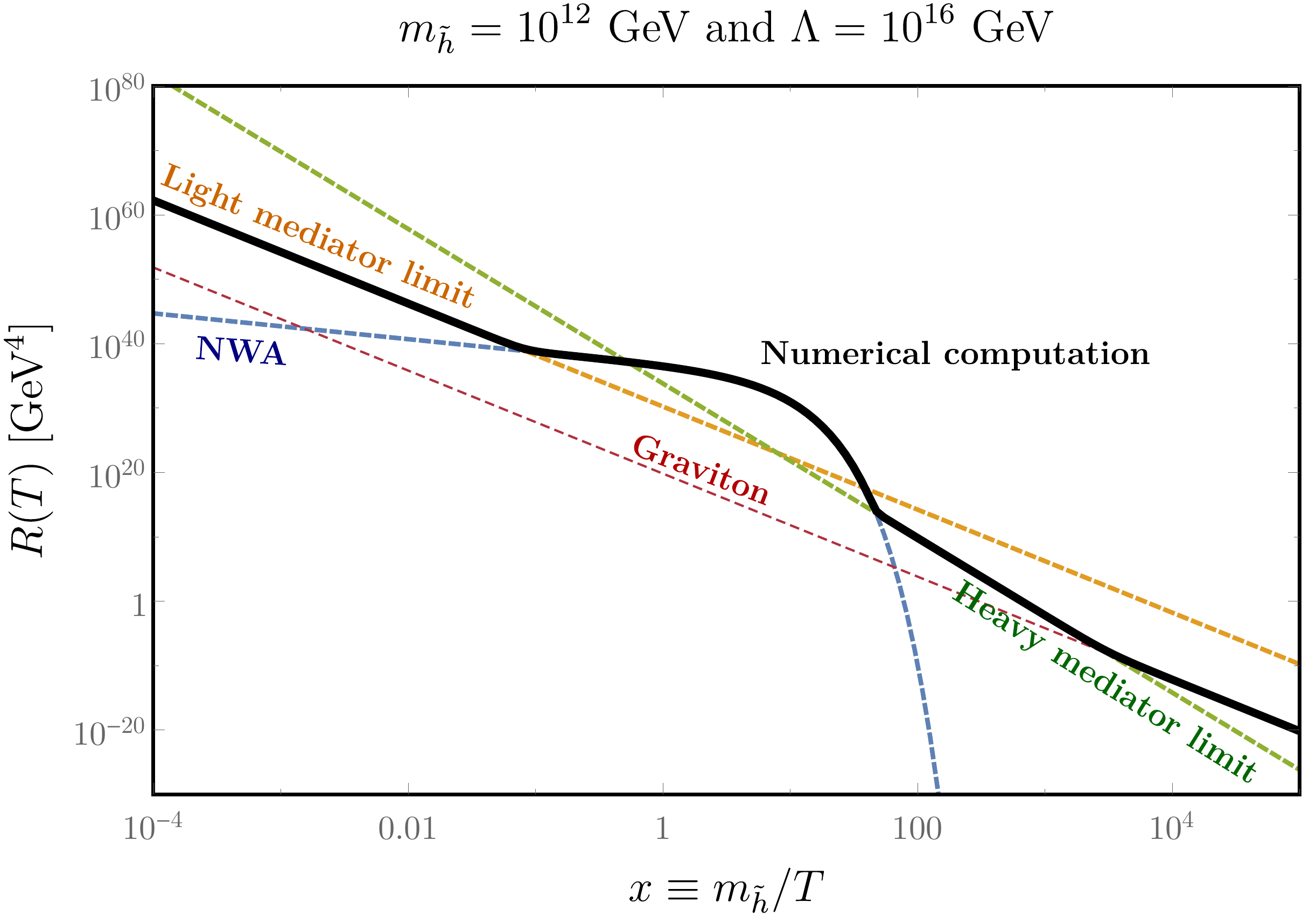}
\caption{Evolution of the production rate $R(T)$ from Eq.(\ref{Eq:rt}) as function of $\mh /T$ for $\Lambda =10^{16} $ GeV and $\mh = 10^{12}$ GeV.}
\label{Fig:rateplot}
\end{figure}

The exact dependence on the temperature of each rate is detailed in the Appendix, and is fundamental in order to understand the behavior of the relic abundance as function of the reheating temperature. For an illustration, we show in Fig.~(\ref{Fig:rateplot}) the production rate $R(T)$ as function 
of the dimensionless parameter $x = \mh /T$ for $\mh=10^{12}$ GeV and $\Lambda=10^{16}$ GeV.

We can clearly distinguish 3 main regimes (in addition, the superheavy mediator regime is seen as
a slight bend in the curve at large $\mh/T$) in this figure. To guide the eye, we have plotted with dashed lines all four regimes as if they were valid at all values of $x$. The line labeled as graviton corresponds to the superheavy mediator with rate proportional to $\alpha$. 
The solid black line corresponds to the full calculation valid for all values of $x$. When $T \gg \mh $, the rate is dominated by the light mediator limit, $R(T)$ decreasing with the temperature at a rate proportional to $T^8$. Then, when the temperature approaches the mass of the mediator, the NWA regime dominates, giving a rate very mildly dependent on the temperature ($R(T) \propto \frac{T}{\mh} K_1(\frac{\mh}{T})$). 
It is only when $T$ drops below $\mh$ that the exponential behavior of the Bessel function dominates. At larger $x$, the rate then drops abruptly to enter in the heavy mediator regime, with a strong dependence on the temperature $R(T)\propto T^{12}$. At still lower temperatures, eventually graviton exchange dominates and the rate again falls as $R\sim T^8$.

In the following subsections, we compute the relic abundance of the dark matter, integrating the production rate in each of these regimes. The integration was made numerically, using the set of equations (\ref{Eq:setboltzmann}), taking into account the effect of non-instantaneous reheating on the relic abundance. However, as it was shown in \cite{Garcia:2017tuj}, the difference induced by the exact non-instantaneous reheating treatment is a multiplicative factor, independent of the model, except when the resonance is important. The analytical expressions in Eq. (\ref{Eq:omega}) are based on the instantaneous reheating approximation and are used only as an aid to describe the results below.

\subsection{Heavy mediator regime}

\begin{figure}[ht!]
\includegraphics[width=0.47\textwidth]{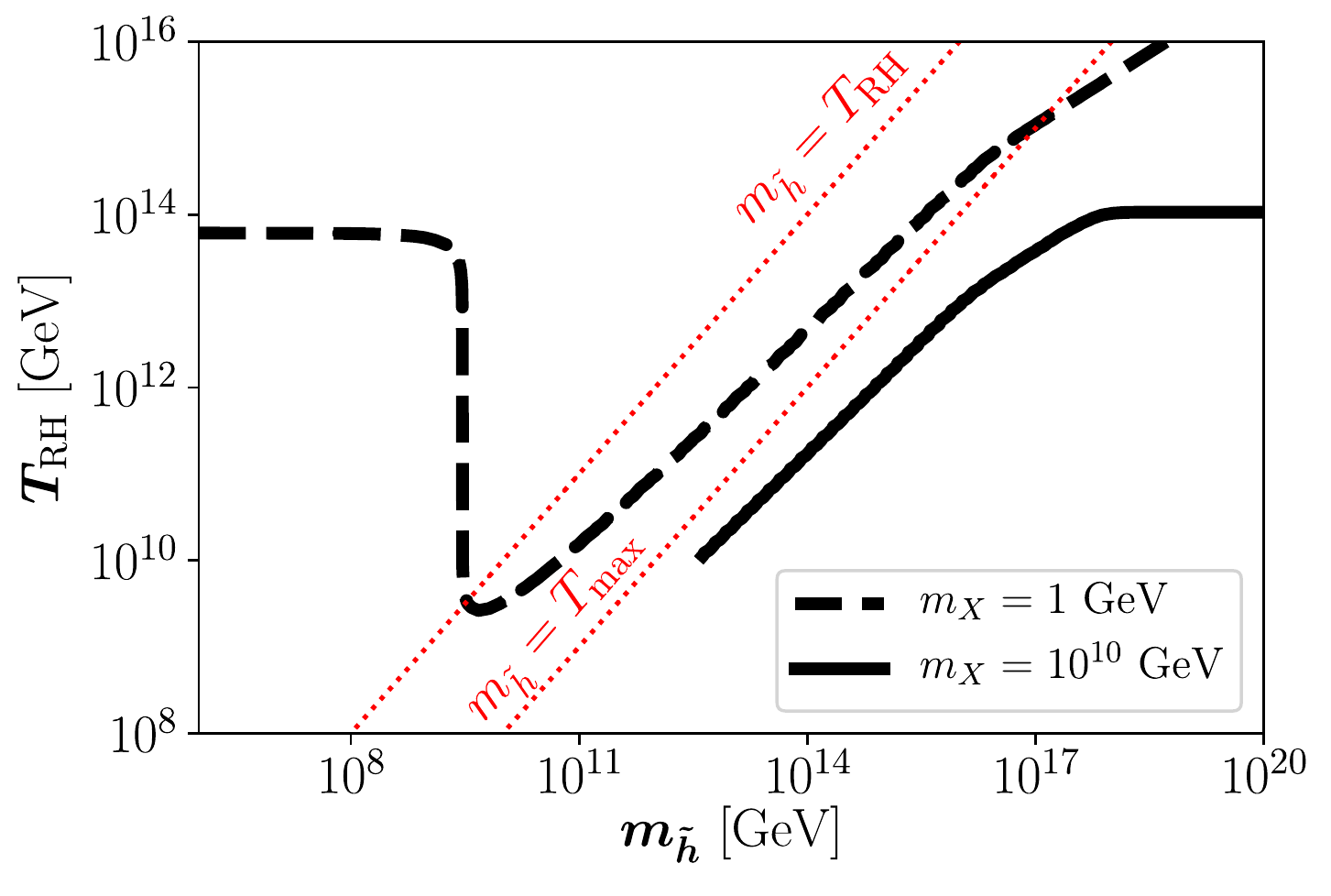}
\caption{Values of $T_\text{RH}$ and $\mh$ giving rise to the observed scalar DM relic abundance for $m_X=1$ GeV and $10^{10}$~GeV, $\gdm=\gsm=1$ and $\Lambda=10^{16}$~GeV.
The dotted diagonal lines with $\mh = T_\text{RH}$ and $\mh = T_\text{\scriptsize{max}} = 100 ~T_\text{RH}$  are shown for reference. 
}
\label{Fig:mtrh}
\end{figure}

In the heavy mediator scenario, for scalar dark matter\footnote{The Appendix also includes the exact formulae for fermionic and vectorial dark matter. However, these differ only by a factor of order one to ten.} one can extract from Eq.~(\ref{Eq:omega}) 
the expression for $\Omega h^2$ in the term
proportional to $\b_3$. This result (based on instantaneous reheating) should be 
multiplied by a ``boost" factor, $B_F$, to account for non-instant reheating. 
It was calculated in \cite{Garcia:2017tuj}
to be $B_F = f(n) \frac{56}{5} \ln \left( \frac{T_\text{max}}{T_\text{RH}} \right)\simeq 20$ for 
$T_\text{max}/T_\text{RH} \sim 100$ and numerically $f(6) \approx 0.4$.
We plot in Fig.~(\ref{Fig:mtrh}) the values of $T_\text{RH}$ and $\mh$ required to obtain
a relic density of $\Omega h^2 \simeq 0.1$ 
for two choices of dark matter masses (1 GeV and $10^{10}$ GeV)  and\footnote{When not specified, we will fix $\gdm=\gsm=1$.} $\Lambda=10^{16}$ GeV. It is important to underline that, to produce this figure, we took into account the enhancement of the production rate due to non-instantaneous reheating. Indeed, as it was shown in \cite{Garcia:2017tuj}, such a high power-law dependence on the reheating temperature implied that the majority of dark matter is produced at the beginning of the reheating process and the approximation of instant reheating is not valid anymore. However, this enhancement {\it does not} depend on the production process of dark matter but only on the ratio $T_\text{\scriptsize{max}}/T_\text{RH}$.

In the heavy mediator regime, we can verify the fact that the relic abundance is compatible with WMAP/Planck data when $m_X \simeq 10^{10}$ GeV and $\Lambda=10^{16} $ GeV, from the analytical expression (\ref{Eq:omega}). Agreement between the curve
and our analytical expression requires the boost factor of about 20.
Thus for $m_{\tilde h}=10^{17}$ GeV, we find $T_\text{RH} \simeq 3 \times 10^{13}$ GeV. 
Note that for this value of $m_X$,
the solid curve cuts off at $T_\text{RH}=10^{10}$ GeV as we must require $T_\text{RH} > m_X$ so that the production of the dark matter is kinematically allowed.    
At lower value of $\mh$ the analytical expression (\ref{Eq:omega}) would require a
significantly larger boost factor as the effect of the pole can not be neglected
and this effect is not accurately taken into account in the analytical expression.
In fact, under close examination of the solid line in Fig.~\ref{Fig:mtrh}, 
we see a change in slope at $\mh \approx 10^{16}$ GeV. At higher masses,
the effect of the pole is safely neglected and the term propotional to $\beta_3$ in Eq. (\ref{Eq:omega}) describes the numerical result reasonably well. 

At still higher $\mh$, the curve flattens out, when the term proportional to $\alpha$ in Eq. (\ref{Eq:omega}) dominates, corresponding to graviton exchange. 
Indeed, Eq.~(\ref{Eq:omega}) shows that when $\mh > 3000 B_F^{1/4} T_\text{RH}$ (for the parameters shown in the figure), graviton exchange dominates and the 
necessary reheat temperature is independent of $\mh$ and is $T_\text{RH} \simeq 10^{14}$ GeV as seen in Fig.~(\ref{Fig:mtrh}).  This is easily understood once one notices that, even if massless, graviton exchange is highly suppressed by Planck mass couplings to the standard-model and dark sector. Note that in the case of graviton exchange there is effectively no boost factor as the rate depends on $T^8$ rather than $T^{12}$.  A large reheating temperature is needed to compensate the weakness of the coupling. Then for all masses $\mh > 7 \times 10^{17}$ GeV, graviton exchange dominates.

For $m_X = 1$ GeV (as seen by the dashed line), the heavy mediator is only important 
at extremely high values of $\mh$ as seen by the slight bend in the curve at
the upper right of the figure. This bend corresponds to the point where
the effect of the pole ceases to dominate as we previously saw for $m_X = 10^{10}$ GeV
and discussed above.

\subsection{Light mediator regime}

As we discussed above, if $m_{\tilde h}$ is lighter than the reheating temperature $T_{RH}$, there is the possibility of resonant production of the mediator $\tilde h_{\mu \nu}$ \cite{blennow_freeze-through_2014}. 
One can easily understand that once the temperature of the thermal bath $T$ dropped to
the value $T\simeq m_{\tilde h}/2$, dark matter production will be enhanced by the rapid s-channel cross section on resonance. 
The important parameter in this case is the width of $\tilde h$. 
Within the narrow width approximation, one can compute the rate and relic density (see the Appendix for details)
which is given in Eq. (\ref{Eq:omega}) by the term proportional to $\beta_2$. 
This expression is obviously independent of $T_{RH}$ because it corresponds to rapid dark matter production around $T\sim \mh$. This pole-phenomena is clearly visible in Fig.~(\ref{Fig:mtrh}), represented by the vertical line (for $m_X=1$ GeV) corresponding to the value $\mh \simeq 5 \times 10^9$ in  good agreement with our analytical computation Eq.~(\ref{Eq:omega}).

For lower values of $\mh$, the pole process occurs at lower temperatures and the dark matter production rate is not sufficient to obtain the correct relic density. The dominant production mode becomes the exchange of the spin-2 mediator offshell. Its contribution is given in Eq. (\ref{Eq:omega}) by the term proportional to $\b_1$. 
In this case, the rate does not scale with the mediator mass and we
expect a specific value of $T_{RH}$ necessary to obtain the correct relic density for a given dark matter mass. We obtain the right amount dark matter with mass, $m_X = 1$ GeV for $T_\text{RH} \simeq 6\times 10^{13}$~GeV, which corresponds to the plateau observed on the left hand side of Fig.~(\ref{Fig:mtrh}).

\subsection{Non-instantaneous reheating}

\begin{figure}[h!]
\centering
\includegraphics[width=0.47\textwidth]{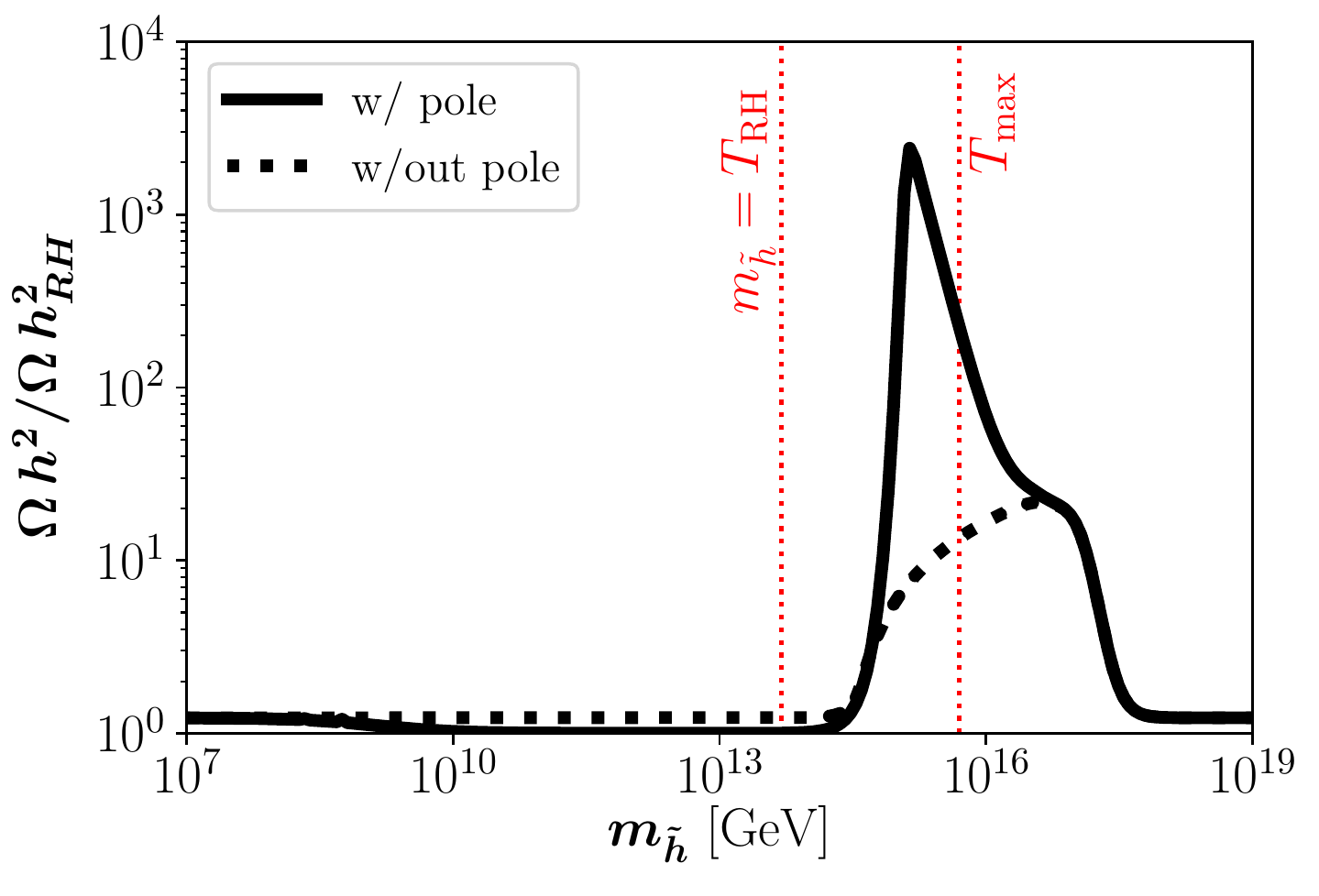}
\caption{Boost factor for scalar dark matter with $m_X = 10^{10}$ GeV due to non-instantaneous reheating as function of $\mh$ for $\Lambda= 10^{16}$ GeV. 
The red dotted-dashed lines correspond to $\mh = T_\text{RH}=5 \times 10^{13}$ GeV and $\mh = T_\text{\scriptsize{max}}= 100~T_\text{RH}$.}
\label{Fig:boost}
\end{figure}

The effects of non-instantaneous reheating on the relic abundance is shown in Fig. (\ref{Fig:boost}), where we plot the ratio of the relic abundance computed with the the exact numerical solution compared to instant-reheating approximation, $\Omega h^2 / \Omega h^2_\text{RH}$. 

There are several effects of non-instantaneous 
reheating, depending on the mass of the mediator $\mh$
relative to $T_\text{RH}$ and $T_\text{\scriptsize{max}}$:

\begin{itemize}
\item{If $\mh \ll T_{RH}$, the dark matter production process is dominated by the exchange of the (light) spin-2 mediator, $\tilde h$. Since the cross section is proportional to $T^2$ ($n=2$), the boost factor $B_F=\Omega h^2 / \Omega h^2|_{\text{RH}}$ is marginal (of the order of 1.5 in accordance with what was found in \cite{Garcia:2017tuj}). This is seen by the horizontal part of the solid line at the smallest values of $\mh$ shown. }

\item{At slightly higher $\mh$, we notice that the boost factor is unity. This is a consequence of the weak influence of the temperature on the production  rate. Indeed, at these values of $\mh$, the rate R(T) is dominated by the pole-production
and  $R(T) \propto T K_1\left(  \frac{\mh}{T} \right)$ as seen in Eq. (\ref{R0b2}). 
If we ignore the pole, then the abundance is characterized by a rate
which is proportional to $T^8$ (as it is for very large $\mh$)
and we obtain a boost factor of $\approx 1.5$ as seen by the dotted line.
 }

\item{If $T_\text{RH} < \mh < T_\text{\scriptsize{max}}$, there is a huge enhancement of the cross section depicted in Fig.~(\ref{Fig:boost}) due to the presence of the pole between $T_\text{RH}$ and $T_\text{\scriptsize{max}}$. This boost factor $B_F$ can even reach a few $\times 10^3$ for this value of $T_\text{RH}$. For lower $T_\text{RH}$, the boost factor can
be much higher as it scales as $\Lambda^2/T_\text{RH}^2$
for a fixed ratio of $\mh/T_\text{RH}$.}

\item{Although  $\mh \gtrsim T_\text{max}$, there is a large boost factor 
associated with the pole, an effect we already saw in the discussion of Fig.~\ref{Fig:mtrh}.
At higher $\mh$,
the boost factor drops from its peak due to the pole,
to the shoulder at around $\mh = 10^{17}$ GeV, where the rate is proportional to 
$T^{12}$ (corresponding to $n=6$ in the cross section due to the exchange of the 
off-shell spin-2 mediator). 
Here, the boost factor is approximately 20 as discussed above. 
The dotted line
ignores the effect of the pole and shows the smooth transition between
rates which vary as $T^8$ to $T^{12}$ to $T^8$. 
}

\item{Finally, at the largest values of $\mh$ shown in the figure,
the production rate is dominated by graviton exchange, and the rate
again varies as $T^8$. The pole can be safely ignored, and the boost 
factor is approximately 1.5 as was found for very low $\mh$.}

\end{itemize}

\section{V. Conclusions}

We have shown that dark matter can naturally be produced through a spin-2 portal. We generalized our study to any massive spin-2 state with stress-energy tensor couplings to the standard model and the dark sector. In a large part of the parameter space, the massive spin-2 portal dominates over the (Planck-suppressed) graviton exchange. We have performed an exhaustive analysis, considering cases where the spin-2 field is both heavier and lighter than the reheating temperature. In both cases, the freeze-in process dominates the production, while enhanced during the reheating phase (heavy mediator case) or through its resonant production (light mediator case). We have also shown that our results are greatly influenced by taking into account the effects of non-instantaneous reheating. Not only do we recover boost factors in the production due to the large dependence on the temperature of the rate $R(T)$, but we have also shown that the presence of a mediator between $T_{RH}$ and $T_{max}$ strongly enhances the relic abundance due to rapid s-channel production when $T \simeq \mh$.  

\vskip.3in
\noindent
{\bf Acknowledgments}

\noindent The authors acknowledge M. Crisostomi for useful discussions. 
This  work was supported by the France-US PICS no. 06482 and PICS MicroDark.
M. Dutra acknowledges support from the Brazilian PhD program ``Ci\^encias sem Fronteiras''-CNPQ Process No. 202055/2015-9.
N. Bernal and Y. Mambrini acknowledge partial support from the European 
 Union Horizon 2020 research and innovation programme under 
 the Marie Sklodowska-Curie: RISE
InvisiblesPlus (grant agreement No 690575)  and 
the ITN Elusives (grant agreement No 674896).
N. Bernal is also supported by the Spanish MINECO under Grant FPA2017-84543-P and by the Universidad Antonio Nariño grant 2017239.
The work of  K.A.O. and M.P. was supported in part
by DOE grant DE--SC0011842 at the University of Minnesota. 

\section*{Appendix}

\section{Spin-2 properties and decay rate}

The propagator of the graviton can be expressed in the Lorentz gauge as  \cite{Giudice:1998ck}
\begin{equation}
 \Pi_{\m\n,\a\b}^{h_0}(k) = \frac{\frac{1}{2}\eta_{\a\n}\eta_{\b\m} + 
\frac{1}{2}\eta_{\a\m}\eta_{\b\n} - \frac{1}{2}\eta_{\a\b}\eta_{\m\n} }{k^2} ~.
\end{equation}
The propagator of the massive state can be written as 
\cite{Giudice:1998ck,Lee:2013bua,Geng:2012hy}
\begin{equation}
\Pi_{\m\n, \a\b}^{h_2}(k) = \frac{i P_{\m\n,\a\b}}{k^2-\mh^2+i\mh \Gh}  ~,
\end{equation}
and the sum over spin polarization states is given by
\begin{equation}
\sum_{s=1}^{5} \varepsilon_{\m\n}(k,s) \varepsilon_{\a\b}(k,s) = 
P_{\m\n,\a\b}~,
\end{equation}
where
\begin{equation}\label{Pmnab}
 P_{\m\n,\a\b} = 
\frac{1}{2}(G_{\m\a}G_{\n\b}+G_{\n\a}G_{\m\b})-\frac{1}{3}G_{\m\n}G_{\a\b},
\end{equation}
with $G_{\m\n} \equiv \eta_{\m\n}-\frac{k_\m k_\n}{\mh^2}$.

The decay modes of the spin-2 state are
\begin{equation}
\G_{\tilde h \to \vp \vp} = N_\vp \frac{g^2_\vp}{960 \pi}\frac{\mh^3}{\L^2} (1-4r_\vp)^{5/2},
\end{equation}
\begin{equation}
\G_{\tilde h \to \psi \psi} = N_\psi \frac{g^2_\psi}{160 \pi}\frac{\mh^3}{\L^2} \left(1+\frac{8}{3}r_\psi\right) (1-4r_\psi)^{3/2},
\end{equation}
and
\begin{equation}
\G_{\tilde h \to V V} = N_V \frac{g^2_V}{960 \pi}\frac{\mh^3}{\L^2}(13+56r_V + 48 r_V^2) (1-4r_V)^{1/2} ,
\end{equation}
where $r_i \equiv m_i^2/\mh^2$. 

The total decay width of the massive spin-2 to SM state is given by :

\begin{equation}
\begin{split}
\Gamma_{\tilde h \rightarrow \text{SM}}= & 4\G_{\tilde h \to \vp \vp}+45\G_{\tilde h \to \psi \psi}+12\G_{\tilde h \to V V} \\ = & \dfrac{43g_\sm^2 m_{\tilde h}^3}{96 \pi \Lambda^2}
\label{Eq:width}
\end{split}
\end{equation}

\section{Amplitudes and rates}

\subsection{Scalar dark matter}

The amplitudes for scalar dark matter with a massless graviton mediator (normal gravity)
are:
\bea
&&
|{\cal M}^0|^2_{h_{\mu \nu}} = \frac{1}{16 M_P^4} \frac{t^2(s+t)^2}{s^2}
\\
&&
|{\cal M}^{1/2}|^2_{h_{\mu \nu}} = \frac{1}{32 M_P^4} \frac{(-t(s+t))(s+2t)^2}{s^2}
\\
&&
|{\cal M}^{1}|^2_{h_{\mu \nu}} = \frac{1}{8 M_P^4} \frac{t^2(s+t)^2}{s^2} \, .
\eea

The corresponding amplitudes assuming a massive spin-2 propagator are:
\bea
&&
|{\cal M}^0|^2_{\tilde h_{\mu \nu}} = \frac{g_{SM}^2 g_{SM}^2}{36 \Lambda^4}\frac{\left[6t(s+t)+s^2\right]^2}{(s-m_{\tilde h}^2)^2 +  \Gamma_{\tilde h}^2m_{\tilde h}^2}
\\
&&
|\M^{1/2}|^2_{\tilde{h}_{\m\n}} = \frac{\gdm^2\gsm^2}{2 \L^4} \frac{(-t(s+t))(s+2t)^2}{(s-\mh^2)^2+\mh^2\Gh^2}
\\
&&
 |\M^1|^2_{\tilde{h}_{\m\n}} = \frac{2\gdm^2\gsm^2}{\L^4} \frac{t^2(s+t)^2}{(s-\mh^2)^2+\mh^2\Gh^2} \, .
\eea

From these amplitudes, we can derive the rate when gravity dominates 
\bea
&&
R^{0}_{h_{\mu \nu}}(T) = \frac{3997\pi^3}{663552000} \frac{T^8}{M^4_P} \equiv \a \frac{T^8}{M^4_P} \, .
\label{R0a}
\eea
In the case of the massive spin state, we distinguish three regimes, depending roughly on the relative value of $\mh$ with respect to $T_\text{RH}$.
\begin{equation}\begin{split}
& R^0_{\tilde h_{\mu \nu}}\Big|_{\mh\ll T}= \frac{\gsm^2 \gdm^2 11351 \pi^3}{124416000}\frac{T^8}{\L^4} \equiv \b_1 \gsm^2 \gdm^2 \frac{T^8}{\L^4} 
\label{R0b1}
\end{split}\end{equation}
\begin{equation}\begin{split}
& R^0_{\tilde h_{\mu \nu}}\Big|_{\mh \sim T} = \frac{\gsm^2 \gdm^2 209}{184 320 \pi^4}\frac{\mh^8}{\Lambda^4} \frac{T}{\Gamma_{\tilde h}} K_1\Big(\frac{\mh}{T}\Big) \\ & \hspace{1.8cm} \equiv \gsm^2 \gdm^2 \b_2 \frac{\mh^8}{\Lambda^4} \frac{T}{\Gamma_{\tilde h}} K_1\Big(\frac{\mh}{T}\Big) \\
& \hspace{1.8cm} = \frac{160\pi \b_2}{(\gdm^2/6 + 215 \gsm^2/3)} \frac{\mh^5 T}{\L^2}K_1\Big(\frac{\mh}{T}\Big)
\label{R0b2}
\end{split}\end{equation}
\begin{equation}
R^0_{\tilde h_{\mu \nu}}\Big|_{\mh\gg T}= \frac{\gsm^2 \gdm^2 205 511 \pi^7}{57153600}\frac{T^{12}}{\L^4 \mh^4} \equiv \gsm^2 \gdm^2 \b_3 \frac{T^{12}}{\L^4 \mh^4}
\label{R0b3}
\end{equation}
In the last line of Eq. \ref{R0b2}, we expressed the width $\Gamma_{\tilde h}$ as function of the parameters of the models (see Eq.\ref{Eq:width}). The values of 
$\alpha$, $\b_1, \b_2$ and $\b_3$ are collected in the Table at the end of this Appendix
which also includes the corresponding coefficients for fermionic and vectorial dark matter.

\subsection{Fermionic dark matter}

\begin{equation}
|\M^0|^2_{h_{\m\n}} = \frac{(-t(s+t))(s+2t)^2}{32 \MP^4 s^2} 
\end{equation}

\begin{equation}
|\M^{1/2}|^2_{h_{\m\n}} = \frac{s^4+10s^3t+42s^2t^2+64st^3+32t^4}{128 \MP^4 s^2} \end{equation}

\begin{equation}
|\M^1|^2_{h_{\m\n}} = \frac{(-t(s+t))(s^2+2t(s+t))}{8 \MP^4 s^2}
\end{equation}

\begin{equation}
|\M^0|^2_{\tilde{h}_{\m\n}} = \frac{\gdm^2\gsm^2}{2 \L^4} \frac{(-t(s+t))(s+2t)^2}{(s-\mh^2)^2+\mh^2\Gh^2}
\end{equation}

\begin{equation}
|\M^{1/2}|^2_{\tilde{h}_{\m\n}} = \frac{\gdm^2\gsm^2}{8 \L^4} \frac{s^4+10s^3t+42s^2t^2+64st^3+32t^4}{(s-\mh^2)^2+\mh^2\Gh^2}
\end{equation}

\begin{equation}
|\M^1|^2_{\tilde{h}_{\m\n}} = \frac{2\gdm^2\gsm^2}{\L^4} \frac{(-t(s+t))(s^2+2t(s+t))}{(s-\mh^2)^2+\mh^2\Gh^2}
\end{equation}
which gives the following rates
\begin{equation}\begin{split}
& R^{1/2}_{ h_{\mu \nu}} = \frac{11351 \pi ^3}{331776000 } \frac{T^8}{\MP^4} 
\end{split}\end{equation}

\begin{equation}
R^{1/2}_{\tilde h_{\mu \nu}}\Big|_{\mh\ll T}= \frac{\gsm^2 \gdm^2 11351 \pi^3}{20736000}\frac{T^8}{\L^4}
\end{equation}

\begin{equation}\begin{split}
& R^{1/2}_{\tilde h_{\mu \nu}}\Big|_{\mh \sim T} = \frac{\gsm^2 \gdm^2 209}{30720 \pi^4}\frac{\mh^8}{\Lambda^4} \frac{T}{\Gamma_{\tilde h}} K_1\Big(\frac{\mh}{T}\Big)
\end{split}\end{equation}

\begin{equation}
R^{1/2}_{\tilde h_{\mu \nu}}\Big|_{\mh\gg T}= \frac{\gsm^2 \gdm^2 205 511 \pi^7}{9525600}\frac{T^{12}}{\L^4 \mh^4} 
\end{equation}

\subsection{Vectorial dark matter}

\begin{equation}
|\M^0|^2_{h_{\m\n}} = \frac{3 t^2(s+t)^2}{16 \MP^4 s^2} 
\end{equation}

\begin{equation}
|\M^{1/2}|^2_{h_{\m\n}} = \frac{(-t(s+t))(5s^2+12t(s+t))}{32 \MP^4 s^2} \end{equation}

\begin{equation}
|\M^1|^2_{h_{\m\n}} = \frac{(s^2+t(s+t))(s^2+3t(s+t))}{8 \MP^4 s^2}
\end{equation}

\begin{equation}
|\M^0|^2_{\tilde{h}_{\m\n}} = \frac{\gdm^2\gsm^2}{36 \L^4} \frac{s^4+12s^3t+120s^2t^2+216st^3+108t^4}{(s-\mh^2)^2+\mh^2\Gh^2}
\end{equation}

\begin{equation}
|\M^{1/2}|^2_{\tilde{h}_{\m\n}} = \frac{\gdm^2\gsm^2}{2 \L^4} \frac{(-t(s+t))(5s^2+12t(s+t))}{(s-\mh^2)^2+\mh^2\Gh^2}
\end{equation}

\begin{equation}
|\M^1|^2_{\tilde{h}_{\m\n}} = \frac{2\gdm^2\gsm^2}{\L^4} \frac{(s^2+t(s+t))(s^2+3t(s+t))}{(s-\mh^2)^2+\mh^2\Gh^2}
\end{equation}
which gives the following rates
\begin{equation}\begin{split}
& R^{1}_{ h_{\mu \nu}} = \frac{5489 \pi ^3}{73728000} \frac{T^8}{\MP^4} 
\end{split}\end{equation}

\begin{equation}
R^{1}_{\tilde h_{\mu \nu}}\Big|_{\mh\ll T}= \frac{\gsm^2 \gdm^2 147563 \pi^3}{124416000}\frac{T^8}{\L^4}
\end{equation}

\begin{equation}\begin{split}
& R^{1}_{\tilde h_{\mu \nu}}\Big|_{\mh \sim T} = \frac{\gsm^2 \gdm^2 2717}{184320 \pi^4}\frac{\mh^8}{\Lambda^4} \frac{T}{\Gamma_{\tilde h}} K_1\Big(\frac{\mh}{T}\Big)
\end{split}\end{equation}

\begin{equation}
R^{1}_{\tilde h_{\mu \nu}}\Big|_{\mh\gg T}= \frac{\gsm^2 \gdm^2 2671643 \pi^7}{57153600}\frac{T^{12}}{\L^4 \mh^4} 
\end{equation}

\begin{table}[ht]
\caption{Coupling coefficients
\label{tab:re}
}
\begin{tabular}{|l|c|c|c|c|}
\hline
 Spin & $\alpha$ & $\beta_1$ & $\beta_2$ & $\beta_3$   \\
\hline
0 & $1.9 \times 10^{-4}$ & $2.8 \times 10^{-3}$ & $1.2 \times 10^{-5}$ & 10.9 \\
\hline
1/2 & $1.1 \times 10^{-3}$ & $1.7 \times 10^{-2}$ & $7.0 \times 10^{-5}$ & 65.2 \\
\hline
1 & $2.3 \times 10^{-3}$ & $3.7 \times 10^{-2}$ & $1.5 \times 10^{-4}$ & 141 \\
\hline
\end{tabular}
\end{table}
\clearpage

\bibliographystyle{apsrev4-1}

\bibliography{spin2}

\begin{thebibliography}{44}%
\makeatletter
\providecommand \@ifxundefined [1]{%
 \@ifx{#1\undefined}
}%
\providecommand \@ifnum [1]{%
 \ifnum #1\expandafter \@firstoftwo
 \else \expandafter \@secondoftwo
 \fi
}%
\providecommand \@ifx [1]{%
 \ifx #1\expandafter \@firstoftwo
 \else \expandafter \@secondoftwo
 \fi
}%
\providecommand \natexlab [1]{#1}%
\providecommand \enquote  [1]{``#1''}%
\providecommand \bibnamefont  [1]{#1}%
\providecommand \bibfnamefont [1]{#1}%
\providecommand \citenamefont [1]{#1}%
\providecommand \href@noop [0]{\@secondoftwo}%
\providecommand \href [0]{\begingroup \@sanitize@url \@href}%
\providecommand \@href[1]{\@@startlink{#1}\@@href}%
\providecommand \@@href[1]{\endgroup#1\@@endlink}%
\providecommand \@sanitize@url [0]{\catcode `\\12\catcode `\$12\catcode
  `\&12\catcode `\#12\catcode `\^12\catcode `\_12\catcode `\%12\relax}%
\providecommand \@@startlink[1]{}%
\providecommand \@@endlink[0]{}%
\providecommand \url  [0]{\begingroup\@sanitize@url \@url }%
\providecommand \@url [1]{\endgroup\@href {#1}{\urlprefix }}%
\providecommand \urlprefix  [0]{URL }%
\providecommand \Eprint [0]{\href }%
\providecommand \doibase [0]{http://dx.doi.org/}%
\providecommand \selectlanguage [0]{\@gobble}%
\providecommand \bibinfo  [0]{\@secondoftwo}%
\providecommand \bibfield  [0]{\@secondoftwo}%
\providecommand \translation [1]{[#1]}%
\providecommand \BibitemOpen [0]{}%
\providecommand \bibitemStop [0]{}%
\providecommand \bibitemNoStop [0]{.\EOS\space}%
\providecommand \EOS [0]{\spacefactor3000\relax}%
\providecommand \BibitemShut  [1]{\csname bibitem#1\endcsname}%
\let\auto@bib@innerbib\@empty
\bibitem [{\citenamefont {Akerib}\ \emph {et~al.}(2017)\citenamefont {Akerib}
  \emph {et~al.}}]{Akerib:2016vxi}%
  \BibitemOpen
  \bibfield  {author} {\bibinfo {author} {\bibfnamefont {D.~S.}\ \bibnamefont
  {Akerib}} \emph {et~al.} (\bibinfo {collaboration} {LUX}),\ }\href {\doibase
  10.1103/PhysRevLett.118.021303} {\bibfield  {journal} {\bibinfo  {journal}
  {Phys. Rev. Lett.}\ }\textbf {\bibinfo {volume} {118}},\ \bibinfo {pages}
  {021303} (\bibinfo {year} {2017})},\ \Eprint
  {http://arxiv.org/abs/1608.07648} {arXiv:1608.07648 [astro-ph.CO]}
  \BibitemShut {NoStop}%
\bibitem [{\citenamefont {Cui}\ \emph {et~al.}(2017)\citenamefont {Cui} \emph
  {et~al.}}]{Cui:2017nnn}%
  \BibitemOpen
  \bibfield  {author} {\bibinfo {author} {\bibfnamefont {X.}~\bibnamefont
  {Cui}} \emph {et~al.} (\bibinfo {collaboration} {PandaX-II}),\ }\href
  {\doibase 10.1103/PhysRevLett.119.181302} {\bibfield  {journal} {\bibinfo
  {journal} {Phys. Rev. Lett.}\ }\textbf {\bibinfo {volume} {119}},\ \bibinfo
  {pages} {181302} (\bibinfo {year} {2017})},\ \Eprint
  {http://arxiv.org/abs/1708.06917} {arXiv:1708.06917 [astro-ph.CO]}
  \BibitemShut {NoStop}%
\bibitem [{\citenamefont {Aprile}\ \emph {et~al.}(2017)\citenamefont {Aprile}
  \emph {et~al.}}]{Aprile:2017iyp}%
  \BibitemOpen
  \bibfield  {author} {\bibinfo {author} {\bibfnamefont {E.}~\bibnamefont
  {Aprile}} \emph {et~al.} (\bibinfo {collaboration} {XENON}),\ }\href
  {\doibase 10.1103/PhysRevLett.119.181301} {\bibfield  {journal} {\bibinfo
  {journal} {Phys. Rev. Lett.}\ }\textbf {\bibinfo {volume} {119}},\ \bibinfo
  {pages} {181301} (\bibinfo {year} {2017})},\ \Eprint
  {http://arxiv.org/abs/1705.06655} {arXiv:1705.06655 [astro-ph.CO]}
  \BibitemShut {NoStop}%
\bibitem [{\citenamefont {Silveira}\ and\ \citenamefont
  {Zee}(1985)}]{Silveira:1985rk}%
  \BibitemOpen
  \bibfield  {author} {\bibinfo {author} {\bibfnamefont {V.}~\bibnamefont
  {Silveira}}\ and\ \bibinfo {author} {\bibfnamefont {A.}~\bibnamefont {Zee}},\
  }\href {\doibase 10.1016/0370-2693(85)90624-0} {\bibfield  {journal}
  {\bibinfo  {journal} {Phys. Lett.}\ }\textbf {\bibinfo {volume} {161B}},\
  \bibinfo {pages} {136} (\bibinfo {year} {1985})}\BibitemShut {NoStop}%
\bibitem [{\citenamefont {McDonald}(1994)}]{McDonald:1993ex}%
  \BibitemOpen
  \bibfield  {author} {\bibinfo {author} {\bibfnamefont {J.}~\bibnamefont
  {McDonald}},\ }\href {\doibase 10.1103/PhysRevD.50.3637} {\bibfield
  {journal} {\bibinfo  {journal} {Phys. Rev.}\ }\textbf {\bibinfo {volume}
  {D50}},\ \bibinfo {pages} {3637} (\bibinfo {year} {1994})},\ \Eprint
  {http://arxiv.org/abs/hep-ph/0702143} {arXiv:hep-ph/0702143 [HEP-PH]}
  \BibitemShut {NoStop}%
\bibitem [{\citenamefont {Burgess}\ \emph {et~al.}(2001)\citenamefont
  {Burgess}, \citenamefont {Pospelov},\ and\ \citenamefont {ter
  Veldhuis}}]{Burgess:2000yq}%
  \BibitemOpen
  \bibfield  {author} {\bibinfo {author} {\bibfnamefont {C.~P.}\ \bibnamefont
  {Burgess}}, \bibinfo {author} {\bibfnamefont {M.}~\bibnamefont {Pospelov}}, \
  and\ \bibinfo {author} {\bibfnamefont {T.}~\bibnamefont {ter Veldhuis}},\
  }\href {\doibase 10.1016/S0550-3213(01)00513-2} {\bibfield  {journal}
  {\bibinfo  {journal} {Nucl. Phys.}\ }\textbf {\bibinfo {volume} {B619}},\
  \bibinfo {pages} {709} (\bibinfo {year} {2001})},\ \Eprint
  {http://arxiv.org/abs/hep-ph/0011335} {arXiv:hep-ph/0011335 [hep-ph]}
  \BibitemShut {NoStop}%
\bibitem [{\citenamefont {Davoudiasl}\ \emph {et~al.}(2005)\citenamefont
  {Davoudiasl}, \citenamefont {Kitano}, \citenamefont {Li},\ and\ \citenamefont
  {Murayama}}]{Davoudiasl:2004be}%
  \BibitemOpen
  \bibfield  {author} {\bibinfo {author} {\bibfnamefont {H.}~\bibnamefont
  {Davoudiasl}}, \bibinfo {author} {\bibfnamefont {R.}~\bibnamefont {Kitano}},
  \bibinfo {author} {\bibfnamefont {T.}~\bibnamefont {Li}}, \ and\ \bibinfo
  {author} {\bibfnamefont {H.}~\bibnamefont {Murayama}},\ }\href {\doibase
  10.1016/j.physletb.2005.01.026} {\bibfield  {journal} {\bibinfo  {journal}
  {Phys. Lett.}\ }\textbf {\bibinfo {volume} {B609}},\ \bibinfo {pages} {117}
  (\bibinfo {year} {2005})},\ \Eprint {http://arxiv.org/abs/hep-ph/0405097}
  {arXiv:hep-ph/0405097 [hep-ph]} \BibitemShut {NoStop}%
\bibitem [{\citenamefont {Djouadi}\ \emph {et~al.}(2012)\citenamefont
  {Djouadi}, \citenamefont {Lebedev}, \citenamefont {Mambrini},\ and\
  \citenamefont {Quevillon}}]{hportal}%
  \BibitemOpen
  \bibfield  {author} {\bibinfo {author} {\bibfnamefont {A.}~\bibnamefont
  {Djouadi}}, \bibinfo {author} {\bibfnamefont {O.}~\bibnamefont {Lebedev}},
  \bibinfo {author} {\bibfnamefont {Y.}~\bibnamefont {Mambrini}}, \ and\
  \bibinfo {author} {\bibfnamefont {J.}~\bibnamefont {Quevillon}},\ }\href
  {\doibase 10.1016/j.physletb.2012.01.062} {\bibfield  {journal} {\bibinfo
  {journal} {Phys. Lett.}\ }\textbf {\bibinfo {volume} {B709}},\ \bibinfo
  {pages} {65} (\bibinfo {year} {2012})},\ \Eprint
  {http://arxiv.org/abs/1112.3299} {arXiv:1112.3299 [hep-ph]} \BibitemShut
  {NoStop}%
\bibitem [{\citenamefont {Han}\ and\ \citenamefont
  {Zheng}(2015)}]{Han:2015hda}%
  \BibitemOpen
  \bibfield  {author} {\bibinfo {author} {\bibfnamefont {H.}~\bibnamefont
  {Han}}\ and\ \bibinfo {author} {\bibfnamefont {S.}~\bibnamefont {Zheng}},\
  }\href {\doibase 10.1007/JHEP12(2015)044} {\bibfield  {journal} {\bibinfo
  {journal} {JHEP}\ }\textbf {\bibinfo {volume} {12}},\ \bibinfo {pages} {044}
  (\bibinfo {year} {2015})},\ \Eprint {http://arxiv.org/abs/1509.01765}
  {arXiv:1509.01765 [hep-ph]} \BibitemShut {NoStop}%
\bibitem [{\citenamefont {Mambrini}\ \emph {et~al.}(2016)\citenamefont
  {Mambrini}, \citenamefont {Nagata}, \citenamefont {Olive},\ and\
  \citenamefont {Zheng}}]{Mambrini:2016dca}%
  \BibitemOpen
  \bibfield  {author} {\bibinfo {author} {\bibfnamefont {Y.}~\bibnamefont
  {Mambrini}}, \bibinfo {author} {\bibfnamefont {N.}~\bibnamefont {Nagata}},
  \bibinfo {author} {\bibfnamefont {K.~A.}\ \bibnamefont {Olive}}, \ and\
  \bibinfo {author} {\bibfnamefont {J.}~\bibnamefont {Zheng}},\ }\href
  {\doibase 10.1103/PhysRevD.93.111703} {\bibfield  {journal} {\bibinfo
  {journal} {Phys. Rev.}\ }\textbf {\bibinfo {volume} {D93}},\ \bibinfo {pages}
  {111703} (\bibinfo {year} {2016})},\ \Eprint
  {http://arxiv.org/abs/1602.05583} {arXiv:1602.05583 [hep-ph]} \BibitemShut
  {NoStop}%
\bibitem [{\citenamefont {Arcadi}\ \emph {et~al.}(2015)\citenamefont {Arcadi},
  \citenamefont {Mambrini},\ and\ \citenamefont {Richard}}]{zportal}%
  \BibitemOpen
  \bibfield  {author} {\bibinfo {author} {\bibfnamefont {G.}~\bibnamefont
  {Arcadi}}, \bibinfo {author} {\bibfnamefont {Y.}~\bibnamefont {Mambrini}}, \
  and\ \bibinfo {author} {\bibfnamefont {F.}~\bibnamefont {Richard}},\ }\href
  {\doibase 10.1088/1475-7516/2015/03/018} {\bibfield  {journal} {\bibinfo
  {journal} {JCAP}\ }\textbf {\bibinfo {volume} {1503}},\ \bibinfo {pages}
  {018} (\bibinfo {year} {2015})},\ \Eprint {http://arxiv.org/abs/1411.2985}
  {arXiv:1411.2985 [hep-ph]} \BibitemShut {NoStop}%
\bibitem [{\citenamefont {Escudero}\ \emph {et~al.}(2016)\citenamefont
  {Escudero}, \citenamefont {Berlin}, \citenamefont {Hooper},\ and\
  \citenamefont {Lin}}]{Escudero:2016gzx}%
  \BibitemOpen
  \bibfield  {author} {\bibinfo {author} {\bibfnamefont {M.}~\bibnamefont
  {Escudero}}, \bibinfo {author} {\bibfnamefont {A.}~\bibnamefont {Berlin}},
  \bibinfo {author} {\bibfnamefont {D.}~\bibnamefont {Hooper}}, \ and\ \bibinfo
  {author} {\bibfnamefont {M.-X.}\ \bibnamefont {Lin}},\ }\href {\doibase
  10.1088/1475-7516/2016/12/029} {\bibfield  {journal} {\bibinfo  {journal}
  {JCAP}\ }\textbf {\bibinfo {volume} {1612}},\ \bibinfo {pages} {029}
  (\bibinfo {year} {2016})},\ \Eprint {http://arxiv.org/abs/1609.09079}
  {arXiv:1609.09079 [hep-ph]} \BibitemShut {NoStop}%
\bibitem [{\citenamefont {Arcadi}\ \emph
  {et~al.}(2017{\natexlab{a}})\citenamefont {Arcadi}, \citenamefont {Ghosh},
  \citenamefont {Mambrini}, \citenamefont {Pierre},\ and\ \citenamefont
  {Queiroz}}]{zpportal}%
  \BibitemOpen
  \bibfield  {author} {\bibinfo {author} {\bibfnamefont {G.}~\bibnamefont
  {Arcadi}}, \bibinfo {author} {\bibfnamefont {P.}~\bibnamefont {Ghosh}},
  \bibinfo {author} {\bibfnamefont {Y.}~\bibnamefont {Mambrini}}, \bibinfo
  {author} {\bibfnamefont {M.}~\bibnamefont {Pierre}}, \ and\ \bibinfo {author}
  {\bibfnamefont {F.~S.}\ \bibnamefont {Queiroz}},\ }\href {\doibase
  10.1088/1475-7516/2017/11/020} {\bibfield  {journal} {\bibinfo  {journal}
  {JCAP}\ }\textbf {\bibinfo {volume} {1711}},\ \bibinfo {pages} {020}
  (\bibinfo {year} {2017}{\natexlab{a}})},\ \Eprint
  {http://arxiv.org/abs/1706.04198} {arXiv:1706.04198 [hep-ph]} \BibitemShut
  {NoStop}%
\bibitem [{\citenamefont {Arcadi}\ \emph
  {et~al.}(2017{\natexlab{b}})\citenamefont {Arcadi}, \citenamefont {Dutra},
  \citenamefont {Ghosh}, \citenamefont {Lindner}, \citenamefont {Mambrini},
  \citenamefont {Pierre}, \citenamefont {Profumo},\ and\ \citenamefont
  {Queiroz}}]{Arcadi:2017kky}%
  \BibitemOpen
  \bibfield  {author} {\bibinfo {author} {\bibfnamefont {G.}~\bibnamefont
  {Arcadi}}, \bibinfo {author} {\bibfnamefont {M.}~\bibnamefont {Dutra}},
  \bibinfo {author} {\bibfnamefont {P.}~\bibnamefont {Ghosh}}, \bibinfo
  {author} {\bibfnamefont {M.}~\bibnamefont {Lindner}}, \bibinfo {author}
  {\bibfnamefont {Y.}~\bibnamefont {Mambrini}}, \bibinfo {author}
  {\bibfnamefont {M.}~\bibnamefont {Pierre}}, \bibinfo {author} {\bibfnamefont
  {S.}~\bibnamefont {Profumo}}, \ and\ \bibinfo {author} {\bibfnamefont
  {F.~S.}\ \bibnamefont {Queiroz}},\ }\href@noop {} {\  (\bibinfo {year}
  {2017}{\natexlab{b}})},\ \Eprint {http://arxiv.org/abs/1703.07364}
  {arXiv:1703.07364 [hep-ph]} \BibitemShut {NoStop}%
\bibitem [{\citenamefont {Mambrini}\ \emph {et~al.}(2013)\citenamefont
  {Mambrini}, \citenamefont {Olive}, \citenamefont {Quevillon},\ and\
  \citenamefont {Zaldivar}}]{so10}%
  \BibitemOpen
  \bibfield  {author} {\bibinfo {author} {\bibfnamefont {Y.}~\bibnamefont
  {Mambrini}}, \bibinfo {author} {\bibfnamefont {K.~A.}\ \bibnamefont {Olive}},
  \bibinfo {author} {\bibfnamefont {J.}~\bibnamefont {Quevillon}}, \ and\
  \bibinfo {author} {\bibfnamefont {B.}~\bibnamefont {Zaldivar}},\ }\href
  {\doibase 10.1103/PhysRevLett.110.241306} {\bibfield  {journal} {\bibinfo
  {journal} {Phys. Rev. Lett.}\ }\textbf {\bibinfo {volume} {110}},\ \bibinfo
  {pages} {241306} (\bibinfo {year} {2013})},\ \Eprint
  {http://arxiv.org/abs/1302.4438} {arXiv:1302.4438 [hep-ph]} \BibitemShut
  {NoStop}%
\bibitem [{\citenamefont {Benakli}\ \emph {et~al.}(2017)\citenamefont
  {Benakli}, \citenamefont {Chen}, \citenamefont {Dudas},\ and\ \citenamefont
  {Mambrini}}]{Benakli:2017whb}%
  \BibitemOpen
  \bibfield  {author} {\bibinfo {author} {\bibfnamefont {K.}~\bibnamefont
  {Benakli}}, \bibinfo {author} {\bibfnamefont {Y.}~\bibnamefont {Chen}},
  \bibinfo {author} {\bibfnamefont {E.}~\bibnamefont {Dudas}}, \ and\ \bibinfo
  {author} {\bibfnamefont {Y.}~\bibnamefont {Mambrini}},\ }\href {\doibase
  10.1103/PhysRevD.95.095002} {\bibfield  {journal} {\bibinfo  {journal} {Phys.
  Rev.}\ }\textbf {\bibinfo {volume} {D95}},\ \bibinfo {pages} {095002}
  (\bibinfo {year} {2017})},\ \Eprint {http://arxiv.org/abs/1701.06574}
  {arXiv:1701.06574 [hep-ph]} \BibitemShut {NoStop}%
\bibitem [{\citenamefont {Dudas}\ \emph {et~al.}(2017)\citenamefont {Dudas},
  \citenamefont {Mambrini},\ and\ \citenamefont {Olive}}]{gravitino}%
  \BibitemOpen
  \bibfield  {author} {\bibinfo {author} {\bibfnamefont {E.}~\bibnamefont
  {Dudas}}, \bibinfo {author} {\bibfnamefont {Y.}~\bibnamefont {Mambrini}}, \
  and\ \bibinfo {author} {\bibfnamefont {K.}~\bibnamefont {Olive}},\ }\href
  {\doibase 10.1103/PhysRevLett.119.051801} {\bibfield  {journal} {\bibinfo
  {journal} {Phys. Rev. Lett.}\ }\textbf {\bibinfo {volume} {119}},\ \bibinfo
  {pages} {051801} (\bibinfo {year} {2017})},\ \Eprint
  {http://arxiv.org/abs/1704.03008} {arXiv:1704.03008 [hep-ph]} \BibitemShut
  {NoStop}%
\bibitem [{\citenamefont {Hinshaw}\ \emph {et~al.}(2013)\citenamefont {Hinshaw}
  \emph {et~al.}}]{Hinshaw:2012aka}%
  \BibitemOpen
  \bibfield  {author} {\bibinfo {author} {\bibfnamefont {G.}~\bibnamefont
  {Hinshaw}} \emph {et~al.} (\bibinfo {collaboration} {WMAP}),\ }\href
  {\doibase 10.1088/0067-0049/208/2/19} {\bibfield  {journal} {\bibinfo
  {journal} {Astrophys. J. Suppl.}\ }\textbf {\bibinfo {volume} {208}},\
  \bibinfo {pages} {19} (\bibinfo {year} {2013})},\ \Eprint
  {http://arxiv.org/abs/1212.5226} {arXiv:1212.5226 [astro-ph.CO]} \BibitemShut
  {NoStop}%
\bibitem [{\citenamefont {Ade}\ \emph {et~al.}(2016)\citenamefont {Ade} \emph
  {et~al.}}]{Ade:2015xua}%
  \BibitemOpen
  \bibfield  {author} {\bibinfo {author} {\bibfnamefont {P.~A.~R.}\
  \bibnamefont {Ade}} \emph {et~al.} (\bibinfo {collaboration} {Planck}),\
  }\href {\doibase 10.1051/0004-6361/201525830} {\bibfield  {journal} {\bibinfo
   {journal} {Astron. Astrophys.}\ }\textbf {\bibinfo {volume} {594}},\
  \bibinfo {pages} {A13} (\bibinfo {year} {2016})},\ \Eprint
  {http://arxiv.org/abs/1502.01589} {arXiv:1502.01589 [astro-ph.CO]}
  \BibitemShut {NoStop}%
\bibitem [{\citenamefont {Nanopoulos}\ \emph {et~al.}(1983)\citenamefont
  {Nanopoulos}, \citenamefont {Olive},\ and\ \citenamefont
  {Srednicki}}]{Nanopoulos:1983up}%
  \BibitemOpen
  \bibfield  {author} {\bibinfo {author} {\bibfnamefont {D.~V.}\ \bibnamefont
  {Nanopoulos}}, \bibinfo {author} {\bibfnamefont {K.~A.}\ \bibnamefont
  {Olive}}, \ and\ \bibinfo {author} {\bibfnamefont {M.}~\bibnamefont
  {Srednicki}},\ }\href {\doibase 10.1016/0370-2693(83)91624-6} {\bibfield
  {journal} {\bibinfo  {journal} {Phys. Lett.}\ }\textbf {\bibinfo {volume}
  {127B}},\ \bibinfo {pages} {30} (\bibinfo {year} {1983})}\BibitemShut
  {NoStop}%
\bibitem [{\citenamefont {Ellis}\ \emph {et~al.}(1984)\citenamefont {Ellis},
  \citenamefont {Hagelin}, \citenamefont {Nanopoulos}, \citenamefont {Olive},\
  and\ \citenamefont {Srednicki}}]{Ellis:1983ew}%
  \BibitemOpen
  \bibfield  {author} {\bibinfo {author} {\bibfnamefont {J.~R.}\ \bibnamefont
  {Ellis}}, \bibinfo {author} {\bibfnamefont {J.~S.}\ \bibnamefont {Hagelin}},
  \bibinfo {author} {\bibfnamefont {D.~V.}\ \bibnamefont {Nanopoulos}},
  \bibinfo {author} {\bibfnamefont {K.~A.}\ \bibnamefont {Olive}}, \ and\
  \bibinfo {author} {\bibfnamefont {M.}~\bibnamefont {Srednicki}},\ }\href
  {\doibase 10.1016/0550-3213(84)90461-9} {\bibfield  {journal} {\bibinfo
  {journal} {Nucl. Phys.}\ }\textbf {\bibinfo {volume} {B238}},\ \bibinfo
  {pages} {453} (\bibinfo {year} {1984})}\BibitemShut {NoStop}%
\bibitem [{\citenamefont {Khlopov}\ and\ \citenamefont
  {Linde}(1984)}]{Khlopov:1984pf}%
  \BibitemOpen
  \bibfield  {author} {\bibinfo {author} {\bibfnamefont {M.~{\relax Yu}.}\
  \bibnamefont {Khlopov}}\ and\ \bibinfo {author} {\bibfnamefont {A.~D.}\
  \bibnamefont {Linde}},\ }\href {\doibase 10.1016/0370-2693(84)91656-3}
  {\bibfield  {journal} {\bibinfo  {journal} {Phys. Lett.}\ }\textbf {\bibinfo
  {volume} {138B}},\ \bibinfo {pages} {265} (\bibinfo {year}
  {1984})}\BibitemShut {NoStop}%
\bibitem [{\citenamefont {Bernal}\ \emph {et~al.}(2017)\citenamefont {Bernal},
  \citenamefont {Heikinheimo}, \citenamefont {Tenkanen}, \citenamefont
  {Tuominen},\ and\ \citenamefont {Vaskonen}}]{fimp}%
  \BibitemOpen
  \bibfield  {author} {\bibinfo {author} {\bibfnamefont {N.}~\bibnamefont
  {Bernal}}, \bibinfo {author} {\bibfnamefont {M.}~\bibnamefont {Heikinheimo}},
  \bibinfo {author} {\bibfnamefont {T.}~\bibnamefont {Tenkanen}}, \bibinfo
  {author} {\bibfnamefont {K.}~\bibnamefont {Tuominen}}, \ and\ \bibinfo
  {author} {\bibfnamefont {V.}~\bibnamefont {Vaskonen}},\ }\href {\doibase
  10.1142/S0217751X1730023X} {\bibfield  {journal} {\bibinfo  {journal} {Int.
  J. Mod. Phys.}\ }\textbf {\bibinfo {volume} {A32}},\ \bibinfo {pages}
  {1730023} (\bibinfo {year} {2017})},\ \Eprint
  {http://arxiv.org/abs/1706.07442} {arXiv:1706.07442 [hep-ph]} \BibitemShut
  {NoStop}%
\bibitem [{\citenamefont {Chung}\ \emph {et~al.}(1999)\citenamefont {Chung},
  \citenamefont {Kolb},\ and\ \citenamefont {Riotto}}]{Chung:1998rq}%
  \BibitemOpen
  \bibfield  {author} {\bibinfo {author} {\bibfnamefont {D.~J.~H.}\
  \bibnamefont {Chung}}, \bibinfo {author} {\bibfnamefont {E.~W.}\ \bibnamefont
  {Kolb}}, \ and\ \bibinfo {author} {\bibfnamefont {A.}~\bibnamefont
  {Riotto}},\ }\href {\doibase 10.1103/PhysRevD.60.063504} {\bibfield
  {journal} {\bibinfo  {journal} {Phys. Rev.}\ }\textbf {\bibinfo {volume}
  {D60}},\ \bibinfo {pages} {063504} (\bibinfo {year} {1999})},\ \Eprint
  {http://arxiv.org/abs/hep-ph/9809453} {arXiv:hep-ph/9809453 [hep-ph]}
  \BibitemShut {NoStop}%
\bibitem [{\citenamefont {Giudice}\ \emph {et~al.}(2001)\citenamefont
  {Giudice}, \citenamefont {Kolb},\ and\ \citenamefont
  {Riotto}}]{Giudice:2000ex}%
  \BibitemOpen
  \bibfield  {author} {\bibinfo {author} {\bibfnamefont {G.~F.}\ \bibnamefont
  {Giudice}}, \bibinfo {author} {\bibfnamefont {E.~W.}\ \bibnamefont {Kolb}}, \
  and\ \bibinfo {author} {\bibfnamefont {A.}~\bibnamefont {Riotto}},\ }\href
  {\doibase 10.1103/PhysRevD.64.023508} {\bibfield  {journal} {\bibinfo
  {journal} {Phys. Rev.}\ }\textbf {\bibinfo {volume} {D64}},\ \bibinfo {pages}
  {023508} (\bibinfo {year} {2001})},\ \Eprint
  {http://arxiv.org/abs/hep-ph/0005123} {arXiv:hep-ph/0005123 [hep-ph]}
  \BibitemShut {NoStop}%
\bibitem [{\citenamefont {Kolb}\ \emph {et~al.}(2003)\citenamefont {Kolb},
  \citenamefont {Notari},\ and\ \citenamefont {Riotto}}]{Kolb:2003ke}%
  \BibitemOpen
  \bibfield  {author} {\bibinfo {author} {\bibfnamefont {E.~W.}\ \bibnamefont
  {Kolb}}, \bibinfo {author} {\bibfnamefont {A.}~\bibnamefont {Notari}}, \ and\
  \bibinfo {author} {\bibfnamefont {A.}~\bibnamefont {Riotto}},\ }\href
  {\doibase 10.1103/PhysRevD.68.123505} {\bibfield  {journal} {\bibinfo
  {journal} {Phys. Rev.}\ }\textbf {\bibinfo {volume} {D68}},\ \bibinfo {pages}
  {123505} (\bibinfo {year} {2003})},\ \Eprint
  {http://arxiv.org/abs/hep-ph/0307241} {arXiv:hep-ph/0307241 [hep-ph]}
  \BibitemShut {NoStop}%
\bibitem [{\citenamefont {Garcia}\ \emph {et~al.}(2017)\citenamefont {Garcia},
  \citenamefont {Mambrini}, \citenamefont {Olive},\ and\ \citenamefont
  {Peloso}}]{Garcia:2017tuj}%
  \BibitemOpen
  \bibfield  {author} {\bibinfo {author} {\bibfnamefont {M.~A.~G.}\
  \bibnamefont {Garcia}}, \bibinfo {author} {\bibfnamefont {Y.}~\bibnamefont
  {Mambrini}}, \bibinfo {author} {\bibfnamefont {K.~A.}\ \bibnamefont {Olive}},
  \ and\ \bibinfo {author} {\bibfnamefont {M.}~\bibnamefont {Peloso}},\ }\href
  {\doibase 10.1103/PhysRevD.96.103510} {\bibfield  {journal} {\bibinfo
  {journal} {Phys. Rev.}\ }\textbf {\bibinfo {volume} {D96}},\ \bibinfo {pages}
  {103510} (\bibinfo {year} {2017})},\ \Eprint
  {http://arxiv.org/abs/1709.01549} {arXiv:1709.01549 [hep-ph]} \BibitemShut
  {NoStop}%
\bibitem [{\citenamefont {Ellis}\ \emph {et~al.}(2016)\citenamefont {Ellis},
  \citenamefont {Garcia}, \citenamefont {Nanopoulos}, \citenamefont {Olive},\
  and\ \citenamefont {Peloso}}]{Ellis:2015jpg}%
  \BibitemOpen
  \bibfield  {author} {\bibinfo {author} {\bibfnamefont {J.}~\bibnamefont
  {Ellis}}, \bibinfo {author} {\bibfnamefont {M.~A.~G.}\ \bibnamefont
  {Garcia}}, \bibinfo {author} {\bibfnamefont {D.~V.}\ \bibnamefont
  {Nanopoulos}}, \bibinfo {author} {\bibfnamefont {K.~A.}\ \bibnamefont
  {Olive}}, \ and\ \bibinfo {author} {\bibfnamefont {M.}~\bibnamefont
  {Peloso}},\ }\href {\doibase 10.1088/1475-7516/2016/03/008} {\bibfield
  {journal} {\bibinfo  {journal} {JCAP}\ }\textbf {\bibinfo {volume} {1603}},\
  \bibinfo {pages} {008} (\bibinfo {year} {2016})},\ \Eprint
  {http://arxiv.org/abs/1512.05701} {arXiv:1512.05701 [astro-ph.CO]}
  \BibitemShut {NoStop}%
\bibitem [{\citenamefont {Garny}\ \emph {et~al.}(2018)\citenamefont {Garny},
  \citenamefont {Palessandro}, \citenamefont {Sandora},\ and\ \citenamefont
  {Sloth}}]{Garny:2017kha}%
  \BibitemOpen
  \bibfield  {author} {\bibinfo {author} {\bibfnamefont {M.}~\bibnamefont
  {Garny}}, \bibinfo {author} {\bibfnamefont {A.}~\bibnamefont {Palessandro}},
  \bibinfo {author} {\bibfnamefont {M.}~\bibnamefont {Sandora}}, \ and\
  \bibinfo {author} {\bibfnamefont {M.~S.}\ \bibnamefont {Sloth}},\ }\href
  {\doibase 10.1088/1475-7516/2018/02/027} {\bibfield  {journal} {\bibinfo
  {journal} {JCAP}\ }\textbf {\bibinfo {volume} {1802}},\ \bibinfo {pages}
  {027} (\bibinfo {year} {2018})},\ \Eprint {http://arxiv.org/abs/1709.09688}
  {arXiv:1709.09688 [hep-ph]} \BibitemShut {NoStop}%
\bibitem [{\citenamefont {Fierz}\ and\ \citenamefont
  {Pauli}(1939)}]{Fierz:1939ix}%
  \BibitemOpen
  \bibfield  {author} {\bibinfo {author} {\bibfnamefont {M.}~\bibnamefont
  {Fierz}}\ and\ \bibinfo {author} {\bibfnamefont {W.}~\bibnamefont {Pauli}},\
  }\href {\doibase 10.1098/rspa.1939.0140} {\bibfield  {journal} {\bibinfo
  {journal} {Proc. Roy. Soc. Lond.}\ }\textbf {\bibinfo {volume} {A173}},\
  \bibinfo {pages} {211} (\bibinfo {year} {1939})}\BibitemShut {NoStop}%
\bibitem [{\citenamefont {Boulware}\ and\ \citenamefont
  {Deser}(1972)}]{Boulware:1973my}%
  \BibitemOpen
  \bibfield  {author} {\bibinfo {author} {\bibfnamefont {D.~G.}\ \bibnamefont
  {Boulware}}\ and\ \bibinfo {author} {\bibfnamefont {S.}~\bibnamefont
  {Deser}},\ }\href {\doibase 10.1103/PhysRevD.6.3368} {\bibfield  {journal}
  {\bibinfo  {journal} {Phys. Rev.}\ }\textbf {\bibinfo {volume} {D6}},\
  \bibinfo {pages} {3368} (\bibinfo {year} {1972})}\BibitemShut {NoStop}%
\bibitem [{\citenamefont {de~Rham}\ \emph {et~al.}(2011)\citenamefont
  {de~Rham}, \citenamefont {Gabadadze},\ and\ \citenamefont
  {Tolley}}]{deRham:2010kj}%
  \BibitemOpen
  \bibfield  {author} {\bibinfo {author} {\bibfnamefont {C.}~\bibnamefont
  {de~Rham}}, \bibinfo {author} {\bibfnamefont {G.}~\bibnamefont {Gabadadze}},
  \ and\ \bibinfo {author} {\bibfnamefont {A.~J.}\ \bibnamefont {Tolley}},\
  }\href {\doibase 10.1103/PhysRevLett.106.231101} {\bibfield  {journal}
  {\bibinfo  {journal} {Phys. Rev. Lett.}\ }\textbf {\bibinfo {volume} {106}},\
  \bibinfo {pages} {231101} (\bibinfo {year} {2011})},\ \Eprint
  {http://arxiv.org/abs/1011.1232} {arXiv:1011.1232 [hep-th]} \BibitemShut
  {NoStop}%
\bibitem [{\citenamefont {Groot~Nibbelink}\ and\ \citenamefont
  {Peloso}(2005)}]{GrootNibbelink:2004hg}%
  \BibitemOpen
  \bibfield  {author} {\bibinfo {author} {\bibfnamefont {S.}~\bibnamefont
  {Groot~Nibbelink}}\ and\ \bibinfo {author} {\bibfnamefont {M.}~\bibnamefont
  {Peloso}},\ }\href {\doibase 10.1088/0264-9381/22/7/008} {\bibfield
  {journal} {\bibinfo  {journal} {Class. Quant. Grav.}\ }\textbf {\bibinfo
  {volume} {22}},\ \bibinfo {pages} {1313} (\bibinfo {year} {2005})},\ \Eprint
  {http://arxiv.org/abs/hep-th/0411184} {arXiv:hep-th/0411184 [hep-th]}
  \BibitemShut {NoStop}%
\bibitem [{\citenamefont {Groot~Nibbelink}\ \emph {et~al.}(2007)\citenamefont
  {Groot~Nibbelink}, \citenamefont {Peloso},\ and\ \citenamefont
  {Sexton}}]{Nibbelink:2006sz}%
  \BibitemOpen
  \bibfield  {author} {\bibinfo {author} {\bibfnamefont {S.}~\bibnamefont
  {Groot~Nibbelink}}, \bibinfo {author} {\bibfnamefont {M.}~\bibnamefont
  {Peloso}}, \ and\ \bibinfo {author} {\bibfnamefont {M.}~\bibnamefont
  {Sexton}},\ }\href {\doibase 10.1140/epjc/s10052-007-0311-x} {\bibfield
  {journal} {\bibinfo  {journal} {Eur. Phys. J.}\ }\textbf {\bibinfo {volume}
  {C51}},\ \bibinfo {pages} {741} (\bibinfo {year} {2007})},\ \Eprint
  {http://arxiv.org/abs/hep-th/0610169} {arXiv:hep-th/0610169 [hep-th]}
  \BibitemShut {NoStop}%
\bibitem [{\citenamefont {Hassan}\ and\ \citenamefont
  {Rosen}(2012)}]{Hassan:2011hr}%
  \BibitemOpen
  \bibfield  {author} {\bibinfo {author} {\bibfnamefont {S.~F.}\ \bibnamefont
  {Hassan}}\ and\ \bibinfo {author} {\bibfnamefont {R.~A.}\ \bibnamefont
  {Rosen}},\ }\href {\doibase 10.1103/PhysRevLett.108.041101} {\bibfield
  {journal} {\bibinfo  {journal} {Phys. Rev. Lett.}\ }\textbf {\bibinfo
  {volume} {108}},\ \bibinfo {pages} {041101} (\bibinfo {year} {2012})},\
  \Eprint {http://arxiv.org/abs/1106.3344} {arXiv:1106.3344 [hep-th]}
  \BibitemShut {NoStop}%
\bibitem [{\citenamefont {Pradler}\ and\ \citenamefont
  {Steffen}(2007)}]{Pradler:2006hh}%
  \BibitemOpen
  \bibfield  {author} {\bibinfo {author} {\bibfnamefont {J.}~\bibnamefont
  {Pradler}}\ and\ \bibinfo {author} {\bibfnamefont {F.~D.}\ \bibnamefont
  {Steffen}},\ }\href {\doibase 10.1016/j.physletb.2007.02.072} {\bibfield
  {journal} {\bibinfo  {journal} {Phys. Lett.}\ }\textbf {\bibinfo {volume}
  {B648}},\ \bibinfo {pages} {224} (\bibinfo {year} {2007})},\ \Eprint
  {http://arxiv.org/abs/hep-ph/0612291} {arXiv:hep-ph/0612291 [hep-ph]}
  \BibitemShut {NoStop}%
\bibitem [{\citenamefont {Rychkov}\ and\ \citenamefont
  {Strumia}(2007)}]{Rychkov:2007uq}%
  \BibitemOpen
  \bibfield  {author} {\bibinfo {author} {\bibfnamefont {V.~S.}\ \bibnamefont
  {Rychkov}}\ and\ \bibinfo {author} {\bibfnamefont {A.}~\bibnamefont
  {Strumia}},\ }\href {\doibase 10.1103/PhysRevD.75.075011} {\bibfield
  {journal} {\bibinfo  {journal} {Phys. Rev.}\ }\textbf {\bibinfo {volume}
  {D75}},\ \bibinfo {pages} {075011} (\bibinfo {year} {2007})},\ \Eprint
  {http://arxiv.org/abs/hep-ph/0701104} {arXiv:hep-ph/0701104 [hep-ph]}
  \BibitemShut {NoStop}%
\bibitem [{\citenamefont {Davidson}\ and\ \citenamefont
  {Sarkar}(2000)}]{Davidson:2000er}%
  \BibitemOpen
  \bibfield  {author} {\bibinfo {author} {\bibfnamefont {S.}~\bibnamefont
  {Davidson}}\ and\ \bibinfo {author} {\bibfnamefont {S.}~\bibnamefont
  {Sarkar}},\ }\href {\doibase 10.1088/1126-6708/2000/11/012} {\bibfield
  {journal} {\bibinfo  {journal} {JHEP}\ }\textbf {\bibinfo {volume} {11}},\
  \bibinfo {pages} {012} (\bibinfo {year} {2000})},\ \Eprint
  {http://arxiv.org/abs/hep-ph/0009078} {arXiv:hep-ph/0009078 [hep-ph]}
  \BibitemShut {NoStop}%
\bibitem [{\citenamefont {Harigaya}\ and\ \citenamefont
  {Mukaida}(2014)}]{Harigaya:2013vwa}%
  \BibitemOpen
  \bibfield  {author} {\bibinfo {author} {\bibfnamefont {K.}~\bibnamefont
  {Harigaya}}\ and\ \bibinfo {author} {\bibfnamefont {K.}~\bibnamefont
  {Mukaida}},\ }\href {\doibase 10.1007/JHEP05(2014)006} {\bibfield  {journal}
  {\bibinfo  {journal} {JHEP}\ }\textbf {\bibinfo {volume} {05}},\ \bibinfo
  {pages} {006} (\bibinfo {year} {2014})},\ \Eprint
  {http://arxiv.org/abs/1312.3097} {arXiv:1312.3097 [hep-ph]} \BibitemShut
  {NoStop}%
\bibitem [{\citenamefont {Mukaida}\ and\ \citenamefont
  {Yamada}(2016)}]{Mukaida:2015ria}%
  \BibitemOpen
  \bibfield  {author} {\bibinfo {author} {\bibfnamefont {K.}~\bibnamefont
  {Mukaida}}\ and\ \bibinfo {author} {\bibfnamefont {M.}~\bibnamefont
  {Yamada}},\ }\href {\doibase 10.1088/1475-7516/2016/02/003} {\bibfield
  {journal} {\bibinfo  {journal} {JCAP}\ }\textbf {\bibinfo {volume} {1602}},\
  \bibinfo {pages} {003} (\bibinfo {year} {2016})},\ \Eprint
  {http://arxiv.org/abs/1506.07661} {arXiv:1506.07661 [hep-ph]} \BibitemShut
  {NoStop}%
\bibitem [{\citenamefont {Blennow}\ \emph {et~al.}(2014)\citenamefont
  {Blennow}, \citenamefont {Fernandez-Martinez},\ and\ \citenamefont
  {Zaldivar}}]{blennow_freeze-through_2014}%
  \BibitemOpen
  \bibfield  {author} {\bibinfo {author} {\bibfnamefont {M.}~\bibnamefont
  {Blennow}}, \bibinfo {author} {\bibfnamefont {E.}~\bibnamefont
  {Fernandez-Martinez}}, \ and\ \bibinfo {author} {\bibfnamefont
  {B.}~\bibnamefont {Zaldivar}},\ }\href {\doibase
  10.1088/1475-7516/2014/01/003} {\bibfield  {journal} {\bibinfo  {journal}
  {Journal of Cosmology and Astroparticle Physics}\ }\textbf {\bibinfo {volume}
  {2014}},\ \bibinfo {pages} {003} (\bibinfo {year} {2014})},\ \bibinfo {note}
  {arXiv: 1309.7348}\BibitemShut {NoStop}%
\bibitem [{\citenamefont {Giudice}\ \emph {et~al.}(1999)\citenamefont
  {Giudice}, \citenamefont {Rattazzi},\ and\ \citenamefont
  {Wells}}]{Giudice:1998ck}%
  \BibitemOpen
  \bibfield  {author} {\bibinfo {author} {\bibfnamefont {G.~F.}\ \bibnamefont
  {Giudice}}, \bibinfo {author} {\bibfnamefont {R.}~\bibnamefont {Rattazzi}}, \
  and\ \bibinfo {author} {\bibfnamefont {J.~D.}\ \bibnamefont {Wells}},\ }\href
  {\doibase 10.1016/S0550-3213(99)00044-9} {\bibfield  {journal} {\bibinfo
  {journal} {Nucl. Phys.}\ }\textbf {\bibinfo {volume} {B544}},\ \bibinfo
  {pages} {3} (\bibinfo {year} {1999})},\ \Eprint
  {http://arxiv.org/abs/hep-ph/9811291} {arXiv:hep-ph/9811291 [hep-ph]}
  \BibitemShut {NoStop}%
\bibitem [{\citenamefont {Lee}\ \emph {et~al.}(2014)\citenamefont {Lee},
  \citenamefont {Park},\ and\ \citenamefont {Sanz}}]{Lee:2013bua}%
  \BibitemOpen
  \bibfield  {author} {\bibinfo {author} {\bibfnamefont {H.~M.}\ \bibnamefont
  {Lee}}, \bibinfo {author} {\bibfnamefont {M.}~\bibnamefont {Park}}, \ and\
  \bibinfo {author} {\bibfnamefont {V.}~\bibnamefont {Sanz}},\ }\href {\doibase
  10.1140/epjc/s10052-014-2715-8} {\bibfield  {journal} {\bibinfo  {journal}
  {Eur. Phys. J.}\ }\textbf {\bibinfo {volume} {C74}},\ \bibinfo {pages} {2715}
  (\bibinfo {year} {2014})},\ \Eprint {http://arxiv.org/abs/1306.4107}
  {arXiv:1306.4107 [hep-ph]} \BibitemShut {NoStop}%
\bibitem [{\citenamefont {Geng}\ \emph {et~al.}(2013)\citenamefont {Geng},
  \citenamefont {Huang}, \citenamefont {Tang},\ and\ \citenamefont
  {Wu}}]{Geng:2012hy}%
  \BibitemOpen
  \bibfield  {author} {\bibinfo {author} {\bibfnamefont {C.-Q.}\ \bibnamefont
  {Geng}}, \bibinfo {author} {\bibfnamefont {D.}~\bibnamefont {Huang}},
  \bibinfo {author} {\bibfnamefont {Y.}~\bibnamefont {Tang}}, \ and\ \bibinfo
  {author} {\bibfnamefont {Y.-L.}\ \bibnamefont {Wu}},\ }\href {\doibase
  10.1016/j.physletb.2013.01.016} {\bibfield  {journal} {\bibinfo  {journal}
  {Phys. Lett.}\ }\textbf {\bibinfo {volume} {B719}},\ \bibinfo {pages} {164}
  (\bibinfo {year} {2013})},\ \Eprint {http://arxiv.org/abs/1210.5103}
  {arXiv:1210.5103 [hep-ph]} \BibitemShut {NoStop}%
\end{thebibliography}%

\end{document}